\documentclass[11pt,preprintnumbers,a4paper,pdflatex,nofootinbib]{revtex4-1}

\usepackage[utf8]{inputenc}
\usepackage{amsmath,amssymb}
\usepackage{physics}
\usepackage{ascmac}
\usepackage{graphicx}
\usepackage[hang,small,bf]{caption}
\usepackage[subrefformat=parens]{subcaption}

\newcommand{\del}{\partial}
\renewcommand{\dd}[1]{\frac{d}{d #1}}
\newcommand{\pp}[1]{\frac{\partial}{\partial #1}}
\newcommand{\Ve}{V_{\mathrm{eff}}}
\newcommand{\dut}{\Longleftrightarrow}
\newcommand{\f}[2]{\frac{#1}{#2}}
\newcommand{\nrb}{\Rightarrow}
\renewcommand{\a}{&}

\newcommand{\m}{$}
\newcommand{\gev}{\,{\rm GeV}}
\newcommand{\tev}{\,{\rm TeV}}

\makeatletter
\renewcommand{\theequation}{%
\thesection.\arabic{equation}}
\@addtoreset{equation}{section}
\makeatother
\begin{document}
\preprint{KUNS-2762}
\title{
Weak scale from Planck scale -- Mass Scale Generation in Classically Conformal Two Scalar System --
}
\author{Junichi Haruna}
\email{j.haruna@gauge.scphys.kyoto-u.ac.jp}
\affiliation{Department of Physics, Kyoto University, Kyoto 606-8502, Japan}
\author{Hikaru Kawai}
\email{hkawai@gauge.scphys.kyoto-u.ac.jp}
\affiliation{Department of Physics, Kyoto University, Kyoto 606-8502, Japan}

\begin{abstract}
In the standard model, the weak scale is the only parameter with mass dimensions.
This means that the standard model itself can not explain the origin of the weak scale.
On the other hand, from the results of recent accelerator experiments, except for some small corrections, the standard model has increased the possibility of being an effective theory up to the Planck scale.
From these facts, it is naturally inferred that the weak scale is determined by some dynamics from the Planck scale.

In order to answer this question, we rely on the multiple point criticality principle as a clue and consider the classically conformal $\mathbb{Z}_2\times \mathbb{Z}_2$ invariant two scalar model as a minimal model in which the weak scale is generated dynamically from the Planck scale.
This model contains only two real scalar fields and does not contain any fermions and gauge fields.

In this model, due to Coleman-Weinberg-like mechanism, one scalar field spontaneously breaks the \m \mathbb{Z}_2\m\, symmetry with a vacuum expectation value connected with the cutoff momentum.
We investigate this using the 1-loop effective potential, renormalization group and large N limit.

We also investigate whether it is possible to reproduce the mass term and vacuum expectation value of the Higgs field by coupling this model with the standard model in the Higgs portal framework.
In this case, the one scalar field that does not break $\mathbb{Z}_2$ can be a candidate for dark matter, and have a mass of about several TeV in appropriate parameters.
On the other hand, the other scalar field breaks $\mathbb{Z}_2$ and has a mass of several tens of GeV.
These results can be verified in near future experiments.
\end{abstract}

\maketitle

\newpage

\tableofcontents
\newpage
	
\section{Intronduction}
\label{s:intro}
The mass of the Higgs particle in the standard model is the only parameter with mass dimensions,
which determines the masses of all particles except for the QCD scale.
In other words, the standard model itself can not explain the origin of the weak scale.

On the other hand, because supersymmetry particles have not been found in recent accelerator experiments such as LHC, the possibility that the standard model is an effective theory up to the Placnk scale has become more realistic except for some small corrections \cite{Buttazzo2013,Hamada2013,Jones2013}.
Therefore, it is natural to think that the weak scale and the Planck scale are related in some kind of dynamics \cite{Landau1955,Terazawa1977a,Terazawa1977b,Terazawa1981}.

Furthermore, the existence of dark matter indicates that some corrections have to be made to the standard model.
The dark matter has been suggested to exist experimentally, and various models have been proposed to explain it, but it has not been identified yet. 

Therefore, it is interesting to consider the case that the modified standard model including dark matter is classically scale invariant (that is, it has no mass scale), and the weak scale is generated by the dynamics of the Higgs field and dark matter. 

As a matter of fact, many people  
\cite{Meissner2007,Foot2008,Iso2009,Iso2009a,Alexander-Nunneley2010,Cline2013,Farzinnia2013,Haba2015,Chataignier2018,Karam2015,Karam2016,Mohamadnejad2019,YaserAyazi2019,YaserAyazi2019a,Heikinheimo2014,Gabrielli2014,Marzola2016}  
have tried to explain the weak scale from its dynamics, considering a classically scale invariant model with added fields corresponding to dark matter.
In many of these works, they introduce new scalar fields, fermions and gauge fields, and some of the scalar fields have vacuum expectation values through the Coleman-Weinberg mechanism 
\cite{Coleman1973}.

Similarly, we first consider the minimal model in which the mass scale is generated dynamically from the cutoff.
In particular, we study the classically conformal $\mathbb{Z}_2\times \mathbb{Z}_2$ invariant two scalar model  \cite{Bornholdt1995,Adams1995}. 
The renormalized Lagrangian of this model is
\begin{align}
\label{e:ltwors}
\mathcal{L}
\a=-\frac{1}{2}(\del_\mu \varphi)^2-\frac{1}{2}(\del_\mu \phi)-V
,\\
\label{e:Vhdn}
V\a=\frac{\rho}{4!} \phi^4 +\frac{\kappa}{4} \phi^2 \varphi^2+\frac{\rho'}{4!}\varphi^4,
\end{align}
where $ \rho, \rho'$\, and $\kappa $ are the renormalized couplings of mass dimension zero.

This Lagrangian is $\mathbb{Z}_2\times \mathbb{Z}_2$ symmetric under the transformation of
\begin{align}
\mathbb{Z}_2:\phi\to-\phi,
\\
\mathbb{Z}_2:\varphi\to -\varphi.
\end{align}
Furthermore, this model is classically scale invariant, with no parameters with mass dimensions.
In the following, we assume $\rho\leq\rho'$ for simplicity.

Here is a comment on classical conformality.
Classical conformality \cite{Bardeen1995} is that Lagrangian is classically scale invariant, that is, it contains only dimensionless parameters.
This can be understood by assuming the multiple point criticality principle (MPP).
MPP is the principle that the parameters of the theory take on the critical value at which the phase of the system changes \cite{Froggatt1996,Froggatt2001,Nielsen2012}.

Let's consider a restriction on the Lagrangian that comes from  MPP.
For example, in the Higgs potential:$V=\lambda \qty|H|^4+\f{m^2}{2}\qty|H|^2$,
the state of $\expval{H}=0$ is stable or metastable for 
 \m m^2\geq0\m\,
 (The symmetry does not break.), and is unstable for
  \m m^2<0\m \,
(The symmetry breaks spontaneously.).
Therefore MPP gives \m m^2 \,\m\,the critical value 0.
If $\,m^2=0$, the Lagrangian has no parameters with mass dimensions and is classically conformal.

In this paper we present the following two points.
First, in the vacuum of this system, $\phi$ spontaneously breaks the $\mathrm{Z}_2$ symmetry (\m\expval{\phi}\neq0\m), while $\varphi$ does not break the $\mathrm{Z}_2$ symmetry (\m \expval{\varphi}=0\m).
The vacuum expectation value \m \expval{\phi}\m\, is generated non-perturbatively\footnote{
Here we use the word 'non-perturbatively' in the sense that it can not be expressed as a power series of the couplings.}
 from the cutoff \m\,\Lambda \m\, and is approximately expressed by Eq.(\ref{e:expvphi}):
\begin{align*}
\tag{\ref{e:expvphi}}
\expval{\phi}^2=\Lambda^2\dfrac{2}{\kappa_0}\exp\qty(-\dfrac{32\pi^2}{3} \dfrac{\rho_0}{\kappa_0^2}).
\end{align*}
Here, $\kappa_0$\, and $ \rho_0 $\, are bare couplings.
The fact that this model is minimal is shown in Section 
\ref{s:phi4} and Section \ref{s:calcepot}.

Secondly, when coupling this model with the standard model, the mechanism examined above may be used as a mechanism to generate the weak scale from the Planck scale.
We study the following model.
The lepton, quark and gauge sectors are the same as the standard model.
On the other hand, the scalar sector is as follows\footnote{
Models similar to this one have already been analyzed in \cite{Farzinnia2013, Sannino2015, Ghorbani2016, Ghorbani2018, Jung2019}.
}
:
\begin{align}
L_{\mathrm{Boson}}\a=-\abs{\del_\mu H}^2-V_{\mathrm{Hh}}+(\text{R.H.S. of Eq.}(\ref{e:ltwors})),
\\
\label{e:VHh}
V_{\mathrm{Hh}}\a=\lambda \abs{H}^4 -\frac{\eta}{2} \phi^2 \abs{H}^2+\frac{\eta'}{2}\varphi^2 \abs{H}^2.
\end{align}
The vacuum expectation value of $\phi$ is generated from the Plank scale, which reproduces the negative mass term of the Higgs field through the second term of Eq.(\ref{e:VHh}).
And we see that the mass $M_\varphi$ of $\varphi$\, is greater than 0.6 TeV, and the mass $M_\phi$ of $\phi$\, is less than about one tenth of $M_\varphi$\,, to reproduce the vacuum expectation value and mass of the Higgs field.
Also, the scalar field $\varphi$ can be regarded as dark matter.

This paper is organized as follows.

In Section \ref{s:phi4} we calculate the effective potential in the simple $\phi^4$ model.
At a first glance, it looks like a vacuum expectation value is generated through the Coleman-Weinberg mechanism.
However, it turns out to be wrong when we improve the effective potential by the renormalization group.

The Section \ref{s:calcepot} is the main part of our paper.
In this section, we calculate the effective potential of the classically conformal $\mathbb{Z}_2\times \mathbb{Z}_2$ invariant two scalar model.
We find that in a wide region of the parameter space (assuming $\rho<\rho'$), one of the scalar fields ($\phi$) have a vacuum expectation value, while the other scalar field ($\varphi$) does not.
The vacuum expectation value $\expval{\phi}$ is generated from the cutoff by  Coleman-Weinberg like mechanism and is cloth to the scale $\mu_*$ where \m\rho(\mu_*)=0\m\,.
Furthermore, unlike the case of the simple $\phi^4$ model in Section \ref{s:phi4}, we observe that this vacuum expectation value does not disappear even if the effective potential is improved by the renormalization group.
Here the relation of the vacuum expectation value of $\phi$ and \m\mu_*\,\m\, is given by Eq.(\ref{e:vev}):
\begin{align*}
\tag{\ref{e:vev}}
\expval{\phi}^2=\f{2}{\kappa(\mu_*)}\mu_*^2.
\end{align*}
Also the masses $M^2_\phi$ and $M^2_\varphi$ of the fields \m\phi\m\, and \m\varphi\m\,, respectively, are given by Eq.(\ref{e:massphik}) and Eq.(\ref{e:massvphi}):
\begin{align*}
\tag{\ref{e:massphik}}
M^2_\phi\a=\frac{\kappa^2(\expval{\phi})}{32\pi^2}\expval{\phi}^2,
\\
\tag{\ref{e:massvphi}}
M^2_\varphi\a=\frac{\kappa(\expval{\phi})}{2}\expval{\phi}^2,
\end{align*}
where \m\kappa(\expval{\phi})\,\m\,  is the value of the running coupling \m\kappa(\mu)\m\, at $\mu=\expval{\phi}$.
Furthermore, we see that Eq.(\ref{e:expvphi}) is derived by solving the renormalization group equation approximately in the region where the coulings are small.

In Section \ref{s:ln}, we calculate the effective potentials of the model corresponding to Section \ref{s:phi4} (\m O(N) \,\m scalar model) and the model corresponding to Section \ref{s:calcepot} (\m O(N)\times O(N)\,\m scalar model), exactly in the large N limit.
Here we confirm the results of Section \ref{s:phi4}, \ref{s:calcepot}.
In particular, we justify that the vacuum expectation value is generated dynamically and the soundness of the theory in which the renormalized couplings are not positive.

In Section \ref{s:param} we consider whether we can explain the weak scale by coupling the classically conformal $\mathbb{Z}_2\times \mathbb{Z}_2$ invariant two scalar model with the standard model.
We calculate the mixing angle between the scalar field \m\phi\m\, and the Higgs field.
We also see that in this model the mass of $\varphi$ must be greater than 0.6 TeV to reproduce the vacuum expectation value and mass of the Higgs field.
On the other hand, the mass of $\phi$ depends on the value $\kappa(v)$, which varies from 0 to about one tenth of $M_\varphi$.
In addition, we show that the scalar field $\varphi$, which is stable because of the \m\mathrm{Z}_2\m\, invariance, can be regarded  as dark matter.

In Section \ref{s:mpp}, we discuss another possibility from MPP.
Specifically, we consider $\mathbb{Z}_2\times \mathbb{Z}_2$ invariant two scalar model for general masses.
When going back to the original MPP, there is a possibility that, besides classical conformality, the parameters of the theory are chosen to be the first order phase transition point.
Then we see that in the cases of the classical conformality and the first-order phase transition point, the vacua are different but the mass scales are of  similar size.

\newpage
\section{No mass generation in simple $\phi^4$ model}
\label{s:phi4}
In this section we calculate the effective potential and see that it looks like a vacuum expectation value is generated through the Coleman-Weinberg mechanism.
However, by improving it by the renormalization group, we see that it is actually an error.
More generally, the symmetry does not break radiatively in a system consisting of one scalar multiplet whose interaction has only one parameter.

The bare Lagrangian of the simple $\phi^4$ model is
\begin{equation}
\mathcal{L}=\f{1}{2}(\del_\mu\phi)^2+\f{m_0^2}{2}\phi^2+\f{\lambda_0}{4!}\phi^4.
\end{equation}

Here \m m_0^2,\,\lambda_0\m\, are bare couplings.
We calculate the 1-loop effective potential of this model according to Coleman-Weinberg\,\cite{Coleman1973}, and the result is
\begin{align}
\Ve&=\f{m(\mu)^2}{2}\phi^2
+\f{\lambda(\mu)}{4!}\phi^4+\f{m_\phi^4}{64\pi^2}\qty(\log\qty(\f{m_\phi^2}{\mu^2})-\f{1}{2}),
\end{align}
where $m^2_\phi:=m^2(\mu)+\f{\lambda(\mu)}{2}\phi^2$.
We have also defined the renormalized couplings $m(\mu),\,\lambda(\mu)$ as follows:
\begin{align}
\begin{cases}
\displaystyle
\f{m_0^2}{\lambda_0}+\f{3}{64\pi^2}\Lambda^2\a= \f{m^2(\mu)}{\lambda(\mu)},\\
\label{e:renl}
\displaystyle
-\lambda_0^{-1}+\f{3}{16\pi^2}\log\qty(\f{\mu}{\Lambda})\a= -\lambda(\mu)^{-1},
\end{cases}
\end{align}
where $\Lambda$ is the cutoff momentum.
We mention here that $\lambda(\mu)$ is positive and increasing monotonically for $0<\mu<\Lambda$ for later use.
From the classical conformality, we require 
\begin{equation*}
\dv[2]{\phi}\Ve\eval{}_{\phi=0}=m=0.
\end{equation*}
Then the effective potential is 
\begin{equation}
\label{e:epot4}
\Ve=\f{\lambda(\mu)}{4!}\phi^4+\f{\lambda^2(\mu)\phi^4}{256\pi^2}\qty(\log\qty(\f{\lambda(\mu)\phi^2}{2\mu^2})-\f{1}{2})
\end{equation}
The minimum condition of this potential is
\begin{equation*}
0=\dd{\phi^2}\Ve=\f{\lambda(\mu)}{12}\phi^2+\f{\lambda^2(\mu)\phi^2}{128\pi^2}\log\qty(\f{\lambda(\mu)\phi^2}{2\mu^2})
=\f{\lambda(\mu)\phi^2}{12}\qty(1+\f{3\lambda(\mu)}{32\pi^2}\log\qty(\f{\lambda(\mu)\phi^2}{2\mu^2})).
\end{equation*}
Then the minimum point is given by
\begin{equation}
\label{e:vev4}
\phi^2=\mu^2\frac{2}{\lambda}\exp(-\f{32\pi^2}{3}\frac{1}{\lambda}).
\end{equation}
Therefore, even in this simple model it seems that $\mathbb{Z}_2$ symmetry breaks spontaneously.
However, this is an error because the above discussion is beyond the scope of perturbation theory.
At the minimum point of \,\m\Ve\m, the contributions of the tree level term (the first term of the Eq.(\ref{e:epot4})) and the quantum correction (the second term of the Eq.(\ref{e:epot4}) are canceled out.
However, because we are now using perturbation theory, Eq.(\ref{e:epot4}) is valid in the region of \m\phi\m\, where the quantum correction is smaller than the tree level term.
Therefore, we can not trust Eq.(\ref{e:vev4}).

This can be clearly seen by improving the effective potential by the renormalization group.
The important thing here is that \m\mu\m\, is arbitrary in Eq.(\ref{e:epot4}).
Therefore, let $\mu=\phi$ so that the quantum correction is always small\footnote{
This operation is called renormalization group improvement  of effective potential.
}
:
\begin{align}
\left.\Ve\right|_{\mu=\phi}&=\Ve(\phi;\lambda(\phi),\phi),\\
&=\frac{\lambda(\phi)}{4!}\phi^4
+\frac{\lambda^2(\phi)\phi^4}{256\pi^2}\qty(\log\qty(\frac{\lambda(\phi)}{2})-\frac{1}{2}).
\end{align}
This expression is valid within the scope of perturbation theory \m(\lambda(\phi) <<1)\m.
Furthermore, if \m\lambda(\phi)<<1\m, the second term is small compared to the first one and can be dropped approximately.
Then we get
\begin{equation}
\label{e:4rgive}
\Ve\simeq\frac{\lambda(\phi)}{4!}\phi^4.
\end{equation}
Because, as mentioned earlier, $\lambda(\phi)$ is positive and increasing monotonically, $\Ve$ is monotonic.
Thus the minimum point of \m\Ve\m\, is \m\phi=0\m .
That is, $\mathbb{Z}_2$ symmetry does not break spontaneously.

\newpage
\section{Mass scale generation in two scalar system}
\label{s:calcepot}
In this section we consider the classically conformal $\mathbb{Z}_2\times \mathbb{Z}_2$ invariant two scalar model. 
The Euclidean bare Lagrangian of this model is given by
\begin{align}
\label{e:lmdl}
\mathcal{L}
=\frac{1}{2}\qty((\del_\mu \varphi)^2+(\del_\mu \phi)^2)+
\frac{1}{2}(m_0^2\phi^2+{m'}_0^{2}\varphi^2)+
\qty(\frac{\rho_0}{4!} \phi^4 +\frac{\kappa_0}{4} \phi^2 \varphi^2+\frac{\rho'_0}{4!}\varphi^4).
\end{align}
Here $m_0,m'_0,\rho_0,\rho'_0,\kappa_0$ are bare couplings.
We consider the case $\rho_0,\rho'_0,\kappa_0>0$, in which the potential is obviously bounded from below\footnote{
The necessary and sufficient condition for the potential to be bounded from below is 
\begin{align}
\rho_0,\rho'_0>0 \text{\, and \,} 3\kappa_0>-\sqrt{\rho_0\rho'_0}.
\end{align}
}.
Now we assume that there is a scale \m \mu_*\m\, in which \m\rho(\mu_*)=0\m\, and \m\rho'(\mu_*)\m, \m\kappa(\mu_*)>0\m , where \,$\,\rho(\mu)$, $\rho'(\mu)$ and $\kappa(\mu)$ are renormalized couplings
\footnote{
This assumption is that $\rho(\mu)$ becomes zero first among the running coupling constants when lowering the renormalization scale.
This is true in a wide range of initial values of the couplings.
} at momentum scale $\mu$
.
Under this assumption, we calculate the effective potential,  imposing the classical conformality.
Then we show that \m\phi\,\m has a nonzero vacuum expectation value by Coleman-Weinberg-like mechanism, unlike in Section \ref{s:phi4}.

Here, $\rho'$ and $\kappa$ are positive, although $\rho(\mu)$ is 0 , so it is natural to think that $\varphi$ does not have a vacuum expectation value.
In the following we calculate the effective potential assuming $\expval{\varphi}=0$ for a while.
We will confirm it in Section \ref{s:cons} and Section \ref{s:ln}.

\subsection{Effective potential and vacuum}
\label{s:epav}
The effective potential for \m\varphi = 0\m\, is calculated as in Section \ref{s:phi4}:
\begin{multline}
\Ve(\phi,\varphi=0)=
\frac{\rho(\mu)}{4!}\phi^4
+\frac{\rho^2(\mu)\phi^4}{256\pi^2}\qty(\log\qty(\frac{\rho(\mu)\phi^2}{2\mu^2})-\frac{1}{2})
+\frac{\kappa^2(\mu)\phi^4}{256\pi^2}\qty(\log\qty(\frac{\kappa(\mu)\phi^2}{2\mu^2})-\frac{1}{2}).
\end{multline}
If we take the renormalization scale $\mu$ to \m\mu_*\m, we have
\begin{align}
\label{e:vemus}
\Ve(\phi,\varphi=0)=\frac{\kappa^2(\mu_*)\phi^4}{256\pi^2}\qty(\log\qty(\frac{\kappa(\mu_*)\phi^2}{2\mu_*^2})-\frac{1}{2}).
\end{align}
Then, the minimum condition of $\Ve$ is 
\begin{align}
0=\dd{\phi^2}\Ve=
\frac{\kappa^2(\mu_*)\phi^2}{128\pi^2}\log\qty(\frac{\kappa(\mu_*)\phi^2}{2\mu_*^2}).
\end{align}
Therefore the minimum point of $\Ve$ is 
\begin{align}
\label{e:vev}
\phi^2=\frac{2}{\kappa(\mu_*)}\mu_*^2.
\end{align}
That is, like Coleman-Weinberg mechanism, $\phi$ has a nonzero vacuum expectation value by quantum correction.
We will discuss in Section \ref{s:vfc} that this vacuum expectation value is related to the cutoff \m\Lambda\m\, non-perturbatively.

It should be noted that the logarithmic term remains small at the minimum point, and the argument is valid within the scope of perturbation theory.
The essential point in the above analysis is that there is a scale \m\mu_*\m\, where \m \rho(\mu_*)=0\m.
We emphasize that such scale does not exist in the simple $\phi^4$ model

\subsection{Mass of \m\phi\,\m\,and \m\varphi\m}
Next, let us find the masses of $\phi$\, and $\varphi$.
Here we denote the vacuum expectation value of $\phi$ as $v$.
That is,
\begin{align}
v^2=\expval{\phi}^2=\f{2}{\kappa(\mu_*)}\mu_*^2.
\end{align}
Because $v$ is $\mu_*$ multiplied by the constant factor $\sqrt{\f{2}{\kappa(\mu_*)}}$, which is not exponentially large, we  can approximate $\kappa(\mu_*)\simeq\kappa(v)$.
Therefore, from Eq.(\ref{e:vemus}),
\begin{align}
\label{e:epotv}
\Ve\simeq \f{\kappa(v)^2\phi^4}{128\pi^2}
\qty (\log\qty(\f{\phi}{v})-\f{1}{4})
\end{align}
Then, the mass $M^2_\phi\,$ of $\phi\,$ including the radiative correction is given by\footnote{
With the 1-loop beta function $\beta_\rho=\frac{3}{16\pi^2}\qty(\rho^2+\kappa^2)$, we can rewrite Eq.(\ref{e:massphik}) as
\begin{align}
M_\phi^2\simeq\frac{\beta_\rho(\mu=v)}{6}v^2,
\end{align}
where we have used $\rho(v)\simeq 0$ at $v\simeq \mu_*$ and $\beta_\rho(\mu=v)\simeq\frac{3\kappa^2(v)}{16\pi^2}$.
}
\begin{align}
\label{e:massphik}
M^2_\phi=\dv[2]{\phi}\Ve\eval{}_{\phi=v}\simeq\frac{\kappa^2(v)}{32\pi^2}v^2.
\end{align}

On the other hand, the mass $M^2_\varphi$ of $\varphi$ is given from the coupling $\f{\kappa}{4}\phi^2\varphi^2$ in Eq.(\ref{e:lmdl}), and therefore we obtain  
\begin{equation}
\label{e:massvphi}
M^2_\varphi
=\frac{\kappa(v)}{2}v^2.
\end{equation}

From (\ref{e:massphik}) and (\ref{e:massvphi}), the mass ratio is 
\begin{equation}
\frac{M^2_\phi}{M^2_\varphi}=\frac{\kappa(v)}{16\pi^2}.
\end{equation}
There is a large difference in the masses of the scalar fields.

\subsection{Relation between vacuum expectation value and cutoff}
\label{s:vfc}
In this section, we explain that the vacuum expectation value given by Eq.(\ref{e:vev}) is generated non-perturbatively from the cutoff\,\m\Lambda\m.
The essential assumption is that $\rho(\mu)$ becomes 0 first among the coupling constants when lowering the renormalization scale.
This assumption holds for a wide range of initial values, and the mechanism is universal.
Here, for simplicity, we consider situations where the renormalization group equation can be solved approximately.

The renormalization group equation for the renormalized couplings is
\begin{align}
\begin{cases}
\mu\dd{\mu}\rho\a=\beta_\rho,\\
\mu\dd{\mu}\rho'\a=\beta_{\rho'},\\
\mu\dd{\mu}\kappa\a=\beta_\kappa,
\end{cases}
\end{align}
where $\beta_\rho,\beta_{\rho'},\beta_\kappa\,$ are the beta functions, which are given in the 1-loop level by \cite{okane2019}:
\begin{equation}
\label{e:beta}
\begin{cases}
\displaystyle
\beta_\rho&=\f{3}{16\pi^2}\qty(\rho^2+\kappa^2),\\
\displaystyle
\beta_\kappa&=\f{1}{16\pi^2}\qty(\rho\kappa+\rho'\kappa+4\kappa^2),\\
\displaystyle
\beta_{\rho'}&=\f{3}{16\pi^2}\qty(\rho'^2+\kappa^2).
\end{cases}
\end{equation}

Here, let's consider the following simple case that
\begin{align}
\begin{cases}
\rho(\mu)\a<<\kappa(\mu),\\
\beta_{\kappa}\a<<1,
\end{cases}
\end{align}
for $\mu_*\leq\mu\leq\Lambda$.
In this case, $\kappa(\mu)\simeq\kappa_0$ and $\beta_\rho\simeq\f{3\kappa^2(\mu)}{16\pi^2}\simeq \f{3\kappa_0^2}{16\pi^2}\,$
hold approximately.
Then the renormalization group equation for $\rho$ can be solved approximately: 
\begin{equation}
\rho(\mu)\simeq\f{3\kappa_0^2}{16\pi^2}\log\qty(\f{\mu}{\Lambda})+\rho_0.
\end{equation}
From this, we obtain 
\begin{align}
0&=\rho(\mu_*)=\f{3\kappa_0^2}{16\pi^2}\log\qty(\f{\mu_*}{\Lambda})+\rho_0,
\end{align}
which gives
\begin{align}
\mu_*&=\Lambda\exp\qty(-\dfrac{16\pi^2}{3} \dfrac{\rho_0}{\kappa_0^2}).
\end{align}
Then, the vacuum expectation value $v^2$ of $\phi$ is given by
\begin{equation}
\label{e:expvphi}
v^2=
\dfrac{2}{\kappa(\mu_*)}\mu_*^2
\simeq \Lambda^2\dfrac{2}{\kappa_0}\exp\qty(-\dfrac{32\pi^2}{3} \dfrac{\rho_0}{\kappa^2_0}).
\end{equation}
Thus the cut off\,\m\Lambda\m\, and the vacuum expectation value are related nonperturbatively.

\subsection{Consistency as a perturbation theory}
\label{s:cons}
In this section, we confirm that the result of Section \ref{s:epav} is consistent as a perturbation theory.
$\Ve$ for the general values of $\phi$ and $\varphi$ is
\begin{multline}
\label{e:generalve}
\Ve(\phi,\varphi)=
\frac{\rho(\mu)}{4!}\phi^4+\frac{\kappa(\mu)}{4}\phi^2\varphi^2+\frac{\rho'(\mu)}{4!}\varphi^4
\\+\frac{m_+^4(\phi,\varphi)}{64\pi^2}\qty(\log\qty(\frac{m_+^2(\phi,\varphi)}{\mu^2})-\frac{1}{2})
+\frac{m_-^4(\phi,\varphi)}{64\pi^2}\qty(\log\qty(\frac{m_-^2(\phi,\varphi)}{\mu^2})-\frac{1}{2}),
\end{multline}
where we have defined
\begin{align}
\label{e:defmpm}
m_{\pm}^2(\phi,\varphi)&
\displaystyle
:=\frac{1}{2}\qty(m_\phi^2+m_\varphi^2\pm\sqrt{(m_\phi^2-m_\varphi^2)^2+4\kappa^2(\mu)\phi^2\varphi^2}),
\\
m_\phi^2(\phi,\varphi)&
\displaystyle
:=\frac{\rho(\mu)}{2}\phi^2+\frac{\kappa(\mu)}{2}\varphi^2,
\\
m_\varphi^2(\phi,\varphi)&
\displaystyle
:=\frac{\kappa(\mu)}{2}\phi^2+\frac{\rho'(\mu)}{2}\varphi^2.
\end{align}

The important thing here is that in the vacuum, $\phi$ and $\varphi$ are massive.
This means that the logarithmic terms $\f{m_\pm^4}{64\pi^2}\log\qty(\f{m^2_\pm}{\mu^2})$ are not large.
Therefore we can trust Eq.(\ref{e:generalve}) or Eq.(\ref{e:vemus}).

From this, it can be understood that $\varphi$ is certainly 0
 around this vacuum as follows.
Let $\mu$ be $\mu_*$ in Eq.(\ref{e:generalve}).
In fact,
because $\kappa$ and $\rho'$ are positive, the quantum correction for them is small.
Therefore, for $\varphi$, it can be regarded as 
\begin{align}
\Ve
\simeq\f{\kappa(\mu_*)}{4}\phi^2\varphi^2+\f{\rho'(\mu_*)}{4!}\varphi^4.
\end{align}
Because $\rho'(\mu_*)$ and $\kappa(\mu_*)$ are positive, $\varphi=0$ is the only minimum point.

We also confirm the validity of the argument in Section \ref{s:epav} by considering the large N limit in the next Section.

\newpage
\section{Large N analysis}
\label{s:ln}
In this section, we justify the results in the previous sections in the large N limit.

\subsection{$O(N)$ scalar model}
\label{s:ons}
Here, we calculate the effective potential of the $O(N)$ scalar model, corresponding to the model of Section \ref{s:phi4} , exactly in the large N limit and examine the vacuum.

The Euclidean bare Lagrangian is 
\begin{equation}
\mathcal{L}=N\qty(\f{1}{2}(\del_{\mu}\phi_{i})^2+\f{m^2_0}{2}\phi_i^2+\frac{\lambda_0}{8}(\phi_i^2)^2),
\end{equation}
where $\phi_i \in \mathbb{R},i=1,\ldots,N,\phi_i^2=\sum_i \phi_i\phi_i\,$, and $m_0,\lambda_0$ are the bare couplings.
Let us calculate the effective potential of this system according to Coleman-Weinberg\footnote{
See \ref{s:calclnve} for detailed calculation.
}.
We consider the path integral\footnote{
We define
$\displaystyle
\int f(x):=\int\,d^4x\,f(x)$.
}:
\begin{align}
\a\int D\phi_i\,\exp\qty[-N\int \f{1}{2}(\del_{\mu}\phi_{i})^2+\f{m^2_0}{2}\phi_i^2+\frac{\lambda_0}{8}(\phi_i^2)^2]
\\
\propto\a\int D\phi_i\,\exp\qty[-N\int \f{1}{2}(\del_{\mu}\phi_{i})^2+
\frac{1}{2\lambda_0}\qty(\f{\lambda_0}{2}\phi_i^2+m_0^2)^2].
\end{align}
Here we introduce the auxiliary field $c(x)\,$ and use the Gauss integral formula:
\begin{align}
\label{e:gauss}
\int Dc\,\exp\qty[-\int \qty(\f{1}{2}c(x)\hat{A}(x)c(x)-c(x)b(x))]
\propto\exp\qty[-\f{1}{2}\tr\log\qty(\hat{A}(x))+\int \f{1}{2}b(x)\hat{A}^{-1}(x)b(x)],
\end{align}
where $\hat{A}$ is a positive definite linear operator.
Then we get
\begin{align}
\a\int D\phi_i\,\exp\qty[-N\int \f{1}{2}(\del_{\mu}\phi_{i})^2+
\frac{1}{2\lambda_0}\qty(\f{\lambda_0}{2}\phi_i^2+m_0^2)^2]
\\
\label{e:aux}
\propto\a \int D\phi_i Dc\,
\exp\qty[-N\int \f{1}{2}(\del_{\mu}\phi_{i})^2
-\frac{1}{2\lambda_0}\qty(c^2-2c\qty(\f{\lambda_0}{2}\phi_i^2+m_0^2))]
,
\end{align}
where the constant factors from the integral of c were ignored.
Now we take the vacuum expectation value of $\phi_i(x)\,$\,  as $\delta_{i,1}\phi\,$\, with  \m O(N)\,\m\, symmetry and 
set $\phi_i(x)=\expval{\phi_i(x)}+\hat{\phi_i}(x)$.

Substituting this into Eq.(\ref{e:aux}) and dropping the first-order term of $\hat{\phi_i}(x)$,we get
\begin{align}
\a\int D\hat{\phi_i}Dc\,
\exp\qty[-N\int 
\f{1}{2}(\del_{\mu}\hat{\phi_{i}})^2+
\f{c}{2}(\hat{\phi_{i}})^2
-\frac{1}{2\lambda_0}\qty(c^2-2c\qty(\f{\lambda_0}{2}\phi^2+m_0^2))]
\\
=\a\int D\hat{\phi_i}Dc\,
\exp\qty[-N\int 
\f{1}{2}\hat{\phi_{i}}\qty(-\del^2+c)\hat{\phi_{i}}
-\frac{1}{2\lambda_0}\qty(c^2-2c\qty(\f{\lambda_0}{2}\phi^2+m_0^2))]
\\
\propto\a\int Dc\,
\exp\qty[-\f{N}{2}\tr\log\qty(-\del^2+c)
-N\int 
-\frac{1}{2\lambda_0}\qty(c^2-2c\qty(\f{\lambda_0}{2}\phi^2+m_0^2))].
\end{align}
We have used the Gauss integral formula (Eq.(\ref{e:gauss})) with $b(x)=0$ when performing the $\hat{\phi_i}$ integral in the last row.

In the large N limit, $c$ integration is simply equivalent to rewriting the integrand to its value at the stationary point for $c$.
That is, 
\begin{align}
\int Dc\,
\a\exp\qty[-\f{N}{2}\tr\log\qty(-\del^2+c)
-N\int 
-\frac{1}{2\lambda_0}\qty(c^2-2c\qty(\f{\lambda_0}{2}\phi^2+m_0^2))]
\\
=
\a\exp\,\qty[-\Gamma(c(\phi),\phi)].
\end{align}
Here we define
\begin{align}
\f{\Gamma(c,\phi)}{N}:=\f{1}{2}\tr\log\qty(-\del^2+c)
+\int 
-\frac{1}{2\lambda_0}\qty(c^2-2c\qty(\f{\lambda_0}{2}\phi^2+m_0^2)),
\end{align}
and $c(\phi)$ is determined by the following equation:
\begin{align}
\label{e:detcg}
\fdv{\Gamma(c,\phi)}{c}\eval{}_{c=c(\phi)}=0.
\end{align}
If c does not depend on $x$, with 
\begin{align}
\tr\log\qty(-\del^2+c)
\a=
\int d^4x\int \f{d^4k}{(2\pi)^4}\, \log(k^2+c
)
\end{align}
we obtain
\begin{align}
\a\f{\Gamma(c,\phi)}{N}\notag\\
\a=
\int d^4x\,\qty{\f{1}{64\pi^2}
\qty[(k^4-c^2)\log(k^2+c)+\f{1}{2}k^4-k^2c]_{k=0}^{k=\Lambda}
-\frac{1}{2\lambda_0}\qty(c^2-2c\qty(\f{\lambda_0}{2}\phi^2+m_0^2))}\\
\a=\int d^4x\,
\qty{\f{1}{64\pi^2}\qty[
(\Lambda^4-c^2)\log(\Lambda^2+c)+c^2\log c+\Lambda^2 c]
-\frac{1}{2\lambda_0}\qty(c^2-2c\qty(\f{\lambda_0}{2}\phi^2+m_0^2))}.
\end{align}
Here $\Lambda$ is the momentum cutoff and we have dropped the constant term.

Furthermore, ignoring the terms that disappear with $\Lambda\to\infty$, we have
\begin{align}
\a\f{\Gamma(c,\phi)}{N}\notag\\
\a=\int d^4x\,
\qty{\f{1}{64\pi^2}\qty[
c^2\qty(\log\qty(\f{c}{\Lambda^2})-\f{1}{2})+2\Lambda^2 c]
-\frac{1}{2\lambda_0}\qty(c^2-2c\qty(\f{\lambda_0}{2}\phi^2+m_0^2))}
\\
\a=\int d^4x\,
\qty{
-\f{c^2}{2\lambda_0}+\f{c^2}{64\pi^2}\qty(\log\qty(\f{c}{\Lambda^2})-\f{1}{2})+c\qty(\f{\phi^2}{2}+\frac{m_0^2}{\lambda_0}+\frac{1}{32\pi^2}\Lambda^2)
}.
\end{align}

Therefore the effective potential $\Ve$ is 
\begin{align}
\f{\Ve}{N}=-\f{c^2}{2\lambda_0}+\f{c^2}{64\pi^2}\qty(\log\qty(\f{c}{\Lambda^2})-\f{1}{2})+c\qty(\f{\phi^2}{2}+\frac{m^2_0}{\lambda_0}+\frac{1}{32\pi^2}\Lambda^2).
\end{align}

Then we define the renormalized couplings \,$\,\lambda(\mu),m(\mu)$ as
\begin{equation}
\begin{cases}
\displaystyle
\f{m_0^2}{\lambda_0}+\frac{1}{32\pi^2}\Lambda^2&
\displaystyle=\f{m^2(\mu)}{\lambda(\mu)},
\\\displaystyle
\label{e:rnlon}
-\frac{1}{\lambda_0}+\frac{1}{32\pi^2}\log\qty(\f{\mu^2}{\Lambda^2})&
\displaystyle
=-\f{1}{\lambda(\mu)}.
\end{cases}
\end{equation}
and the final expression of the effective potential is
\begin{equation}
\label{e:epot}
\f{\Ve}{N}(c(\phi),\phi)=-\f{c^2}{2\lambda(\mu)}+\f{c^2}{64\pi^2}\qty(\log\qty(\f{c}{\mu^2})-\f{1}{2})+c\qty(\f{\phi^2}{2}+\frac{m^2(\mu)}{\lambda(\mu)}),
\end{equation}
where $c(\phi)$ is determined from 
\begin{align}
0&=\fdv{\Gamma(c,\phi)}{c}\eval{}_{c=c(\phi)}
=\del_c\Ve(c,\phi)\eval{}_{c=c(\phi)}
\\
\label{e:detc}
&=-\f{c}{\lambda(\mu)}+\f{c}{32\pi^2}\log\qty(\frac{c}{\mu^2})+\qty(\f{\phi^2}{2}+\f{m^2(\mu)}{\lambda(\mu)})\eval{}_{c=c(\phi)}.
\end{align}

Now let us find the minimum value of $\Ve$.
The extremum condition is
\begin{equation}
0=\del_{\phi}\Ve(c(\phi),\phi)
=\qty(\pdv{c}{\phi} \pdv{c}+\widetilde{\del_{\phi}})\Ve(c(\phi),\phi),
\end{equation}
where $\widetilde{\del_{\phi}}$ means to differentiate the part that depends on $\phi$ explicitly.
Because from the definition of $c(\phi)\,$,
$\pdv{c}\Ve(c(\phi),\phi)=0$ holds for any $\phi$, the extreme condition is
\begin{align}
0=\widetilde{\del_{\phi}}\Ve(C(\phi),\phi)=\f{1}{2}c\phi.
\end{align}
Therefore, we solve Eq.(\ref{e:detc}) in the cases of
\m
(1)\, c=0,\phi\neq0\quad
\m
and
\m
(2)\,\phi=0
\quad
\m
and compare the value of $\Ve$ to find the minimum value.

\renewcommand{\labelenumi}{(\arabic{enumi})}
\begin{enumerate}
\item 
If $c=0$, Eq.(\ref{e:detc}) becomes
\begin{align}
\f{\phi^2}{2}+\f{m^2(\mu)}{\lambda(\mu)}=0.
\end{align}
The solution of this equation is
\begin{align}
\phi^2=
\begin{cases}
\quad\text{no solution}\a\quad(m^2\geq0),\\
\displaystyle
\f{2\qty|m^2(\mu)|}{\lambda(\mu)}\a\quad(m^2<0).
\end{cases}
\end{align}

\item 
If $\phi=0$, Eq.(\ref{e:detc}) becomes
\begin{align}
\label{e:csol}
-\f{c}{\lambda(\mu)}+\a\f{c}{32\pi^2}\log\qty(\frac{c}{\mu^2})+\f{m^2(\mu)}{\lambda(\mu)}=0,
\end{align}
The solution of this equation is
\footnote{
$W$ is the branch of Lambert's W function, where $W\geq -1$.
}
\begin{align}
c=
\begin{cases}
\a\mu_\lambda^2\exp(W\qty(-\f{32\pi^2}{\lambda(\mu)}\f{m^2(\mu)}{\mu_\lambda^2}))
=m^2(\mu)\qty(1+\f{\lambda}{32\pi^2}\log\qty(\f{m^2(\mu)}{\mu^2})+O(\lambda^2(\mu)))\quad(m^2\geq0),
\\
\a\text{no solution}\quad (m^2<0).
\end{cases}
\end{align}
Here we have introuduced $\mu_\lambda^2:=\mu^2\exp\qty(\f{32\pi^2}{\lambda(\mu)}).$

In Eq.(\ref{e:csol}) , there is another solution $c=c_2:$
\footnote{
$W_{-1}$ is the branch of Lambert's W function, where $W\leq -1$.
}
\begin{align}
c_2\a=\mu_\lambda^2\exp(W_{-1}\qty(-\f{32\pi^2}{\lambda(\mu)}\f{m^2(\mu)}{\mu_\lambda^2}))
\\
\label{e:clp}
\a=\mu^2\exp\qty(\f{32\pi^2}{\lambda(\mu)}) -\f{32\pi^2m^2(\mu)}{\lambda(\mu)}+O\qty(m^2(\mu)\f{m^2(\mu)}{\lambda^2(\mu)\mu_\lambda^2}).
\end{align}
However, $c_2$ is a very large value because
\begin{align}
\mu^2\exp\qty(\f{32\pi^2}{\lambda(\mu)})=\Lambda^2\exp\qty(\f{32\pi^2}{\lambda_0})>>\Lambda^2,
\end{align}
from Eq.(\ref{e:rnlon}).
This value corresponds to the Landau pole, where the quantum field theory is not defined.
Therefore $c_2$ is a non-physical solution, so we exclude it from the analysis.
\end{enumerate}
\renewcommand{\labelenumi}{\arabic{enumi}.}

In summary, the minimum point of $\Ve$ is given by
\begin{align}
(\phi^2,c)=
\begin{cases}
\displaystyle
\qty(0,m^2(\mu)+O(\lambda(\mu)))\quad\a(m^2\geq0),\\
\displaystyle
\qty(2\f{\qty|m^2(\mu)|}{\lambda(\mu)},0)\quad\a(m^2<0).
\end{cases}
\end{align}

Thus, we see that \m O(N)\m\:  symmetry does not break when $m^2\geq0$, and breaks spontaneously when $m^2<0$.

Here is a comment on the physical meaning of $c$.
In the above result, $c\simeq m^2(\mu)\, $(the mass of $N\,\phi$’ \,s) when the $O(N)$ symmetry is not broken, and $c=0$ (the mass of $N-1\:$ NG bosons) when the $O(N)$ symmetry is broken.
Therefore, $c$ corresponds to the mass of particles.

\subsection{\m O(N)\times O(N)\m\,scalar model}
\label{s:velnv}
In this section, we calculate the effective potential of the \m O(N)\times O(N)\m\,  scalar model, corresponding to the model in Section \ref{s:calcepot}, exactly in the large N limit, and show that  only one of the scalar fields has a nonzero vacuum expectation value.

The Euclidian bare Lagrangian is 
\begin{multline}
\mathcal{L}
=N\left(\frac{1}{2}(\del_\mu\phi_i)^2+
\frac{1}{2}(\del_\mu\varphi_i)^2+\right.\\
\left.\frac{1}{2}(m_0^2\phi_i^2+{m'}_0^{2}\varphi^2)+
\frac{\rho_0}{8}(\phi_i^2)^2+
\frac{\kappa_0}{4}(\phi_i^2)(\varphi_j^2)+
\frac{\rho'_0}{8}(\varphi_i^2)^2\right),
\end{multline}
where $\phi_i,\varphi_i \in \mathbb{R},\,i=1,\ldots,N,\,\phi_i^2=\sum_i \phi_i\phi_i,\,\varphi_i^2=\sum_i \varphi_i\varphi_i$.

Now, let us consider the case where the bare couplings satisfy $\rho_0'>\rho_0>0\,,\,\kappa_0>0\,\text{\,and\,}\,\rho_0\rho'_0-\kappa_0^2<0.$
In this case, the potential is bounded from below obviously.

We consider the effective potential of this Lagrangian.
The effective potential $\Ve$ is\footnote{
See \ref{s:calclnve} for a detailed calculation.
}
\begin{equation}
\label{e:veln}
\frac{\Ve}{N}=-\frac{1}{2} C^{t}\lambda^{-1} C
+\frac{c^2}{64\pi^2}\qty(\log\frac{c}{\mu^2}-\frac{1}{2})
+\frac{c'^2}{64\pi^2}\qty(\log\frac{c'}{\mu^2}-\frac{1}{2})
+\frac{1}{2}C^{t}\mqty(\phi^2 \\ \varphi^2),
\end{equation}
where $\lambda(\mu):=\mqty(\rho(\mu)&\kappa(\mu)\\ \kappa(\mu)&\rho'(\mu))$ are the renormalized couplings defined by
\begin{align}
-\lambda_0^{-1}+\frac{1}{32\pi^2}\log\qty(\f{\mu^2}{\Lambda^2})\mqty(1\a0\\0\a1)=-\lambda(\mu)^{-1},
\end{align}
Here we have defined $\lambda_0:=\mqty(\rho_0&\kappa_0\\ \kappa_0&\rho'_0)$, and the cutoff and the renormalization scale as $\Lambda$ and \m\mu\m, respectively.
 $m_0^2$ and ${m'}_0^2$ are determined from classical conformality:
\begin{align}
\lambda_0^{-1}\mqty(m_0^2\\ {m'}_0^{2})+\frac{\Lambda^2}{16\pi^2}\mqty(1\\1)=0.
\end{align}

The auxiliary fields \m C^t=\mqty(c(\phi,\varphi)\a c'(\phi,\varphi))^t\m have the meaning of the renormalized mass of each field.
They must be nonnegative and are determined from $\pp{C}\Ve=0.$

In other words, separating $C$ dependence and explicit $\phi,\varphi$ dependence, we write $\Ve$ as $\Ve(C,\phi,\varphi)$,
 then $C=C(\phi,\varphi)$ is determined to satisfy 
\begin{align}
\pp{C}\Ve(C,\phi,\varphi)\eval{}_{C=C(\phi,\varphi)}=0,
\end{align}
that is
\begin{equation}
-\lambda^{-1}C+\frac{1}{32\pi^2}\mqty(c\log\frac{c}{\mu^2}\\ c'\log\frac{c'}{\mu^2})+\frac{1}{2}\mqty(\phi^2 \\ \varphi^2)=0.
\end{equation}
From this, we have
\begin{equation}
\label{e:Ccond}
\begin{cases}
\displaystyle
\frac{1}{2}\phi^2=\bar{\kappa}c'-\frac{c}{32\pi^2}\log\qty(\frac{c}{\mu_{\rho'}^2}),
\\
\displaystyle
\frac{1}{2}\varphi^2=\bar{\kappa}c-\frac{c'}{32\pi^2}\log\qty(\frac{c'}{\mu_{\rho}^2}).
\end{cases}
\end{equation}
Here, we have defined 
\begin{align}
\mqty(\bar{\rho'}&-\bar{\kappa}\\-\bar{\kappa}&\bar{\rho}):=-\lambda^{-1}(\mu)
=\f{1}{\qty|\det \lambda(\mu)|}\mqty(\rho'&-\kappa\\-\kappa&\rho),
\end{align}
and $\mu_{\rho}^2:=\mu^2\exp\qty(-32\pi^2\bar{\rho}),\,\mu_{\rho'}^2:=\mu^2\exp\qty(-32\pi^2\bar{\rho'}).$
Note that  $\bar{\rho'}>\bar{\rho}$ because we assume that $\rho'>\rho$.

Let us find the minimum value of the effective potential.
The extremum condition is 
\begin{equation}
0=\mqty(\del_{\phi}\\\del_{\varphi})\Ve(C(\phi,\varphi),\phi,\varphi)
=\mqty(\pdv{C}{\phi} \pdv{C}+\widetilde{\del_{\phi}} \\\pdv{C}{\varphi} \pdv{C}+\widetilde{\del_{\varphi}})\Ve(C(\phi,\varphi),\phi,\varphi).
\end{equation}
However, according to the definition of $C$,
$\pdv{C}\Ve(C(\phi,\varphi),\phi,\varphi)=0$ holds for any 
$\phi\,\text{and}\,\,\varphi$, so the extreme condition is 
\begin{align}
0=\mqty(\widetilde{\del_{\phi}} \\ \widetilde{\del_{\varphi}})\Ve(C(\phi,\varphi),\phi,\varphi)=\f{1}{2}\mqty(c\phi\\c'\varphi)
\end{align}
Therefore, the minimum value can be found by examining the points of \m
(1)\, c=c'=0\quad
(2)\, c=\varphi=0,c'\neq0\quad
(3)\,\phi=c'=0,c\neq0\quad
(4)\,\phi=\varphi=0,c\neq0,c'\neq0
\,\m\,
 and comparing the extreme values there.
 
In the case of (4), $c$ and $c'$ are very large compared to the cutoff.
These values correspond to the Landau pole and must be regarded as non-physical, as in Section \ref{s:ons}.
Therefore, we exclude it from the analysis.

Substituting these conditions into Eq.(\ref{e:Ccond}) shows 
\begin{align*}
&(1)\quad
c=c'=\phi=\varphi=0
&&\Rightarrow 
&&\Ve=0\\
&(2)\quad
c=\varphi=0,c'=\mu_{\rho}^2,\phi^2=2\bar{\kappa}\mu_{\rho}^2
&&\nrb 
&&\Ve=-\f{\mu^4}{128\pi^2}\exp(-128\pi^2\bar{\rho}-1)\\
&(3)\quad
c'=\phi=0,c=\mu_{\rho'}^2,\varphi^2=2\bar{\kappa}\mu_{\rho'}^2
&&\nrb 
&&\Ve=-\f{\mu^4}{128\pi^2}\exp(-128\pi\bar{\rho'}-1)
\end{align*}
Therefore, because $\bar{\rho}'>\bar{\rho}$, the point of (2)
\begin{align}
\phi^2=2\bar{\kappa}\mu_{\rho}^2\,,\,\varphi^2=0,
\end{align}
 is the vacuum.
From this, it is shown that $O (N)\,$ symmetry of $\phi$ is spontaneously broken, but 
$O(N)\,$ symmetry of $\varphi$ is not broken.

\newpage

\section{Two scalar model as Higgs portal dark matter}
\label{s:param}
In this section, we couple the classically conformal $\mathbb{Z}_2\times \mathbb{Z}_2$ invariant two scalar model with the standard model and investigate whether the mechanism examined in Section \ref{s:calcepot} can be used as a mechanism to generate the weak scale from the Planck scale.
For that purpose , we first discuss how to incorporate it into the standard model in Section \ref{s:cpltosm}.
Next we calculate the mixing angle between $\phi\,$ and the Higgs field in Section  \ref{s:mixang} and see that it is so small that this model is not excluded experimentally.
We also discuss the masses of \m\phi\m\,and\,\m\varphi\m.

\subsection{Coupling to Standard Model}
\label{s:cpltosm}
We consider how to incorporate classically conformal $\mathbb{Z}_2\times \mathbb{Z}_2$ invariant two scalar model into the standard model.
Because \m\phi\,\,\m has a vacuum expectation value while 
\m\varphi\,\,\m does not, it seems that \m\phi\,\,\m may be regarded as the Higgs field and \m\varphi\,\,\m as an unknown scalar field (for example, dark matter).
But a little calculation shows that unfortunately there is no such possibility.

Let us explain this reason.
The mass of Higgs particles is observed to be 125 \m\gev\m and the vacuum expectation value is 246 \m\gev\m.
Therefore, if \m\phi\,\,\m is regarded as the Higgs field, then $M_\phi=125\gev,v=246\gev$, and the ratio is 
\begin{align}
\label{e:rhgs}
\f{M_\phi^2}{v^2}=\qty(\f{125}{246})^2=2.58\times 10^{-1}.
\end{align}
However, if \m \kappa(v)\,\m is at most 1 in Eq.(\ref{e:massphik}), then 
\begin{align}
\f{M_\phi^2}{v^2}=\frac{\kappa^2(v)}{32\pi^2}<\frac{1}{32\pi^2}=3.16\times 10^{-3}<<2.58\times 10^{-1}
\end{align}
holds, which contradicts Eq.(\ref{e:rhgs}).

Therefore, when incorporating this model into the standard model, $\phi\,$ and $\varphi\,$ must be unknown real scalar fields different from the Higgs field.
As a minimal model that satisfies this condition, we consider
\footnote{
As mentioned in Section \ref{s:intro}, models similar to this one have been analyzed in
\cite{Farzinnia2013,Sannino2015,Ghorbani2016,Ghorbani2018,Jung2019}.
}
\begin{align}
\label{e:msm}
L\a=-\abs{\del_\mu H}^2-V_{\mathrm{Hh}}+(\text{R.H.S. of Eq.}(\ref{e:lmdl})).
\\
V_{\mathrm{Hh}}\a
=\lambda \abs{H}^4 -\frac{\eta}{2} \phi^2 \abs{H}^2+\frac{\eta'}{2}\varphi^2 \abs{H}^2.
\end{align}

Here \m H\,\,\m is the Higgs doublet of the standard model, and the coupling constants \,$\,\eta,\eta'$ between the Higgs field and $\phi\text{\,\,or\,\,}\varphi$ are assumed to be sufficiently small $(\eta,\eta'<<1)$ for the mechanism considered in Section \ref{s:calcepot} to hold.

In this model, \m\phi\,\,\m has a vacuum expectation value generated from the Planck scale, which gives the negative mass  term of the Higgs field through the Higgs portal coupling($\eta$).
That is, ignoring the mixing, if
$\,\eta\,$ satisfies
\begin{align}
\label{e:etacond}
\eta \expval{\phi}^2\simeq m_H^2(=(125\gev)^2),
\end{align}
then the Higgs field feels the potential of
\begin{align}
\lambda \abs{H}^4 -\frac{\eta}{2} \expval{\phi}^2 \abs{H}^2
\simeq 
\lambda \abs{H}^4 -\frac{m_H^2}{2} \abs{H}^2,
\end{align}
so we can reproduce the spontaneous symmetry breaking in the weak scale (\m \expval{H}=\f{m_H}{2\sqrt{\lambda}}\m).

We emphasize that in this model the weak scale \m \expval{H}\m is generated indirectly but non-perturbatively from the Planck scale \m \Lambda\m, as
\begin{align*}
\text{Planck scale}\,\Lambda \to \expval{\phi}\to
\text{weak scale}\,\expval{H}
\end{align*}

\subsection{Mixing angle between \m\phi\,\m\, and Higgs field}
\label{s:mixang}
In this section, we consider the mixing of \m\phi\,\m\, and $H$ and calculate the mixing angle.
In addition, we obtain the restrictions on the masses of \m\phi\m\, and \m\varphi\,\,\m from the conditions that this model reproduces the vacuum expectation value and mass of the Higgs field.
Note that although the mass of \m\phi\,\m\, changes from Eq.(\ref{e:massphik}) due to the mixing, the mass of \m\varphi\,\m\, remains as Eq.(\ref{e:massvphi}).

From Eq.(\ref{e:epotv}), the potential including the loop effect of $\varphi$  is 
\begin{equation}
V_{\phi H}:=
\lambda\abs{H}^4-\f{\eta}{2}\phi^2\abs{H}^2+\frac{\kappa^2}{128\pi^2}\phi^4
\qty (\log\qty(\f{\phi}{v})-\f{1}{4}).
\end{equation}
If we take the unitary gauge and set $H=\f{1}{\sqrt{2}}\mqty(0\\h)$, $V_{\phi H}$ becomes
\begin{align}
V_{\phi H}=\f{\lambda}{4}h^4-\f{\eta}{4}\phi^2h^2+\frac{\kappa^2}{128\pi^2}\phi^4
\qty (\log\qty(\f{\phi}{v})-\f{1}{4}).
\end{align}
Here, we denote the vacuum expectation values of $\phi$ and $h$ as 
$v_0$ and $v_H$, respectively.
These are determined by
\begin{align}
\label{e:vevpsm}
0\a=\pdv{\phi}V_{\phi H}\eval{}_{\phi=v_0,H=v_H}=\frac{\kappa^2}{32\pi^2}v_0^3
\log\qty(\f{v_0}{v})-\f{\eta}{2}v_0v_H^2,
\\
\label{e:vevhsm}
0\a=\pdv{h}V_{\phi H}\eval{}_{\phi=v_0,h=v_H}=\lambda v_H^3-\f{\eta}{2}v_0^2v_H.
\end{align}
From Eq.(\ref{e:etacond}) and Eq.(\ref{e:vevpsm}), we get
\begin{align}
v_0=v\qty(1+
\qty(\f{4\pi v_Hm_H}{\kappa v^2})^2
+O\qty(
\qty(\f{4\pi v_Hm_H}{\kappa v^2})^4)).
\end{align}
Here, from Eq.(\ref{e:massvphi}),
\begin{align}
\qty(\f{4\pi v_Hm_H}{\kappa v^2})^2
=\qty(\f{2\pi v_Hm_H}{M_\varphi^2})^2 \simeq \qty(\f{M_\varphi}{0.44 \tev})^{-4}
\end{align}
But this value is small if $M_\varphi$ is at the TeV scale, as we will see later.
Therefore, for simplicity we set $v_0=v$ in the following argument. 
Then $v_H$ satisfies
\begin{align}
\label{e:vevh}
v_H^2=\f{\eta}{2\lambda}v^2,
\end{align}
from Eq.(\ref{e:vevhsm}).

Let us calculate the mixing angle.
The quadratic terms of $\hat{h}$ and $\hat{\phi}$ in the potential  $V_{\phi H}$ are
\begin{equation}
\f{1}{2}\qty(\qty(3\lambda v_H^2-\f{\eta}{2} v^2)\hat{h}^2-2\eta v v_H\hat{h}\hat{\phi}+
\qty(M_\phi^2-\f{\eta v_H^2}{2})\hat{\phi}^2)
=\f{1}{2}\mqty(\hat{h}&\hat{\phi})M^2\mqty(\hat{h}\\\hat{\phi}).
\end{equation}
Here, $M^2$ is the mass matrix defined by
\begin{align}
\label{e:M2}
M^2:=\mqty(3\lambda v_H^2-\f{\eta}{2} v^2&-\eta v v_H\\-\eta v v_H&M_\phi^2-\f{\eta v_H^2}{2})
=m_h^2\mqty(1&-\f{v_H}{v}\\-\f{v_H}{v}&\f{M_\phi^2}{m_h^2}-\f{1}{2}\qty(\f{v_H}{v})^2),
\end{align}
where we have introduced\footnote{
Note that $m_h$ is not the Higgs field's mass \m m_H=125\,\gev\,\m  itself.
} 
$m_h^2=\eta v^2$ and used Eq.(\ref{e:vevh}).
The eigenvalues of $M^2\,$ gives the masses of the observed particles.

Let us examine the restrictions on the masses of $\phi\,$ and $\varphi$\, from the conditions for the mass squares to be positive.
These conditions are 
\begin{align}
\label{e:detm}
\det M^2\a>0,\\
\label{e:trm}
\tr M^2\a>0.
\end{align}

From Eq.(\ref{e:detm}) , Eq.(\ref{e:massphik}) and Eq.(\ref{e:massvphi}), we get the restrictions of 
\begin{align}
0\a<\det M^2=\f{M_\phi^2}{m_h^2}-\f{3}{2}\qty(\f{v_H}{v})^2,
\\
\label{e:vcondk}
\a\dut \f{3}{2}(m_hv_H)^2<M_\phi^2 v^2=\frac{\kappa^2(v)}{32\pi^2}v^4=\f{1}{8\pi^2}\qty(M_\varphi^2)^2,
\\\a\dut
\label{e:condc}
M_\varphi^2>2\sqrt{3}\pi(m_hv_H)\simeq (579\gev)^2.
\end{align}
From Eq.(\ref{e:vcondk}), we also have
\begin{align}
\label{e:vcond}
v^2>\f{4\sqrt{3}}{\kappa}\pi(m_hv_H)\gtrsim\f{(819 \gev)^2}{\kappa}
\end{align}

On the other hand, Eq.(\ref{e:trm}) becomes
\begin{align}
\label{e:trmk}
0<1+\f{M_\phi^2}{m_h^2}-\f{1}{2}\qty(\f{v_H}{v})^2
\dut v>v_H \sqrt{\f{1}{2}\qty(1+\f{M_\phi^2}{m_h^2})},
\end{align}
which is satisfied if $\kappa$ is not so large\footnote{
In detail, by solving Eq.(\ref{e:trmk}) for $\kappa$\,, we get
\begin{align}
\label{e:condk}
32\pi^2\qty(\f{2m_h^2}{v_H^2}-\f{m^2_h}{v^2})>\kappa.
\end{align}
However, for example, if we assume that $\kappa\lesssim10$, $v^2$ is limited to
\begin{align}
v^2\gtrsim (259 \gev)^2,
\end{align}
from Eq.(\ref{e:vcond}).
Then
\begin{align}
32\pi^2\qty(\f{2m_h^2}{v_H^2}-\f{m^2_h}{v^2})\gtrsim32\pi^2\qty(2\qty(\f{125}{246})^2-\qty(\f{125}{259})^2)>89>10\gtrsim \kappa.
\end{align}
Therofore, Eq.(\ref{e:condk}) is trivially satisfied.
}.

Now, we define $b$ and $c$ as
\begin{align}
M^2=m_h^2\mqty(1&-\f{v_H}{v}\\-\f{v_H}{v}&\qty(\f{v_H}{v})^2\qty(\f{M_\phi^2v^2}{m_h^2v_H^2}-\f{1}{2}))
=:m_h^2\mqty(1\a-b\\-b\a c).
\end{align}
That is, 
\begin{align}
b\a=\f{v_H}{v}=\qty(\f{v}{246\,\gev})^{-1},
\\
c\a=\qty(\f{v_H}{v})^2\qty(\f{M_\phi^2v^2}{m_h^2v_H^2}-\f{1}{2})
\simeq\qty(\f{v}{246\,\gev})^{-2}\qty(\f{3}{2}\qty(\f{M_\varphi}{579\,\gev})^4-\f{1}{2}).
\end{align}
Here we have used $M_\phi^2 v^2=8\pi^2 M_\varphi^4$, which is obtained from Eq.(\ref{e:massphik}) and Eq.(\ref{e:massvphi}), and $m_h\simeq m_H =125\gev$.

If we denote the eigenvalues of this matrix as $r\,m_h^2\,$ , $r\,$ satisfy
\begin{align}
(r-1)(r-c)-b^2=0,
\end{align}
which gives
\begin{align}
r=\f{c+1\pm \sqrt{(c-1)^2+4b^2}}{2}=:r_\pm
\end{align}
The mixing angle $\theta\,$ is given by 
\begin{align}
\theta\a:=\arctan\qty(\f{1-r_+}{b})
=\arctan \qty(\f{2b}{c-1+\sqrt{(c-1)^2+4b^2}})\\
\a\simeq \f{b}{c-1}\quad\qty(\text{if\,\,}\,\qty|\f{b}{c-1}|<<1.).
\end{align}

\begin{figure}[h]
\centering
\includegraphics[width=14cm,pagebox=cropbox]{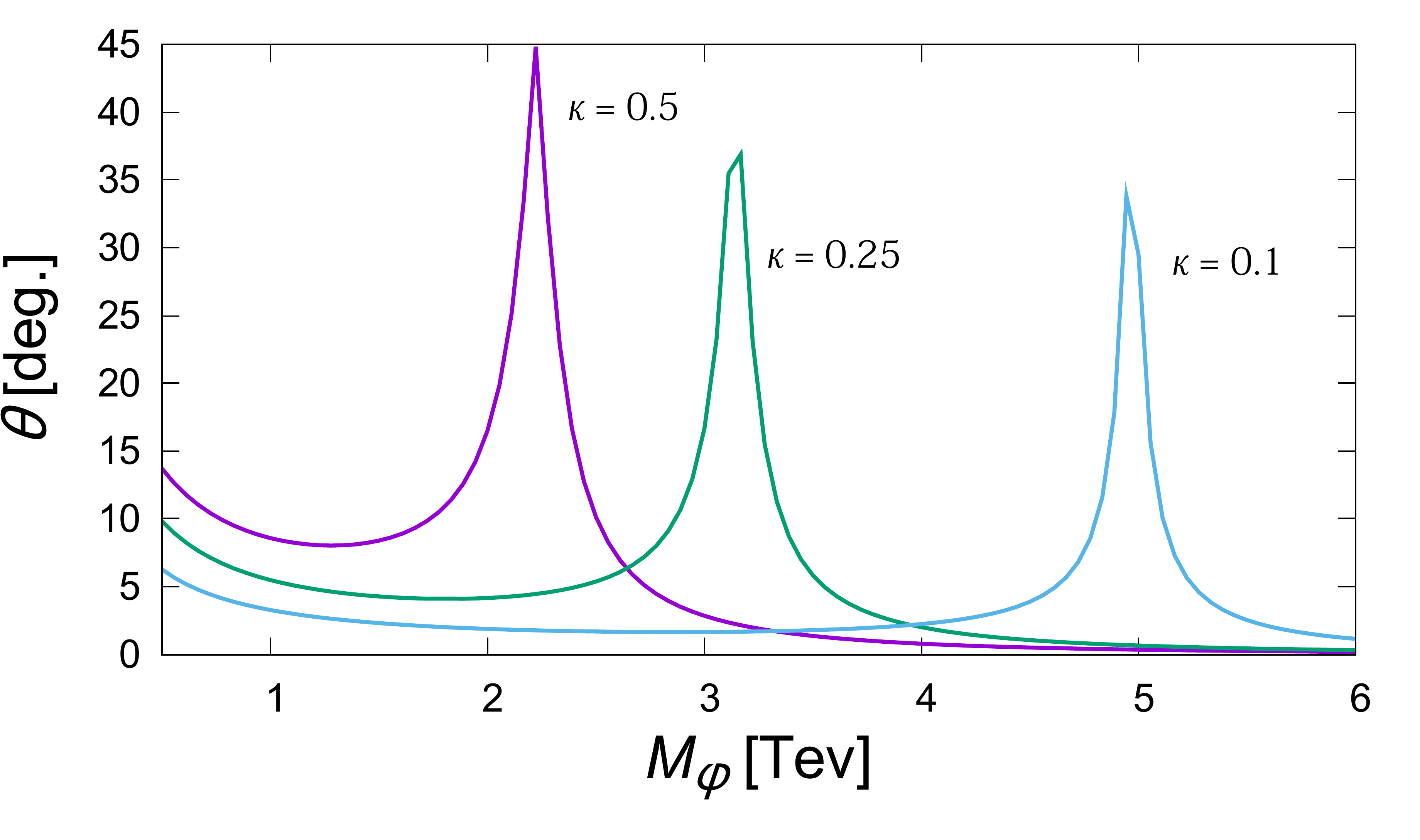}
\caption{the mixing angle against $M_\varphi$ for $\kappa=0.5,0.25,0.1$}
\label{f:ma}
\end{figure}

Let us summarize the results of this section:
\begin{align}
M_\varphi\a=\sqrt{\f{\kappa}{2}}\,v,\\
M_\phi\a=\f{\kappa}{4\sqrt{2}\pi}\,v\a\a=\f{\sqrt{\kappa}}{12.6}\,M_\varphi,\\
\theta \a=\arctan \qty(\f{2b}{c-1+\sqrt{(c-1)^2+4b^2}})
\a\a\simeq \f{b}{c-1}\quad\qty(\text{if\,\,}\,\qty|\f{b}{c-1}|<<1.),\\
b\a=\f{v_H}{v}
\a\a=\sqrt{\kappa}\,\f{0.174}{M_\varphi\,/\tev},\\
c\a=\qty(\f{v_H}{v})^2\qty(\f{M_\phi^2v^2}{m_h^2v_H^2}-\f{1}{2})
\a\a=\kappa\,3.36\times 10^{-3}\qty(\qty(\f{M_\varphi}{\tev})^2-\f{4.47}{(M_\varphi\,/\tev)^2}).
\end{align}
Fig.\ref{f:ma} depicts the mixing angle as a function of $M_\varphi$\, with $\kappa$\, fixed.
The peak appearing in this figure corresponds to the value of  $M_\varphi$ for which $c = 1$.
As can be seen in Fig.\ref{f:ma}, if $M_\varphi$ are lighter or heavier than the value of the peak position, the mixing angle is sufficiently small for each $\kappa$.

As an example of a lighter mass case,
if $\kappa=0.1,M_\varphi=1\,\tev,$ then $M_\phi=25\,\gev,v=4.47\,\tev,\,\theta=5.7\times 10^{-2}.$
On the other hand, as an example of a heavier mass case,
if $\kappa=0.1,M_\varphi=6\,\tev,$ then $M_\phi=151\,\gev,v=26.8\,\tev,\,\theta=2.0\times 10^{-2}$

Because $\varphi $ has mass of the TeV scale and couples to the Higgs field, it can be regarded as the Higgs portal scalar dark matter \cite{Hamada2014,Hamada2017}.
From the experiments, the mass of the Higgs portal scalar dark matter is limited to be greater than 0.7$\pm$0.2 GeV  \cite{Hamada2017,Cui2017}.
That is,
\begin{align}
\label{e:vmassbnd}
\bar{M}_\varphi>0.7\pm 0.2\,\, \tev,
\end{align}
where $\bar{M}^2_\varphi$ is the mass of $\varphi$ that includes the contribution from the Higgs field:
\begin{align}
\bar{M}^2_\varphi:=M^2_\varphi+\f{\eta'}{2}v_H^2.
\end{align}
Eq.(\ref{e:vmassbnd}) is consistent with Eq.(\ref{e:condc}).
Also we have the constraint on the coupling constant between the Higgs field and the dark matter $\varphi$, which comes from the thermal abundance of dark matter:
\begin{align}
\bar{M}_\varphi\simeq\eta' \times3.3\, \tev,
\end{align}
which fixes the coupling $\eta'$.

In addition to this, it is interesting to consider restrictions on the mass $\bar{M}_\varphi$ of $\varphi$ when making other assumptions.
For example, if we assume the Higgs inflation \cite{Hamada2014a,Hamada2014b,Bezrukov2014,Hamada2015},
 from the tensor-to-scalar ratio of the cosmic microwave background \cite{Ade2016}, there is an upper bound on $\bar{M}_\varphi:$
\begin{align}
\bar{M}_\varphi<1.1\,\,\tev.
\end{align}

On the other hand, we can not identify $\phi$ here, which has the smaller mass.
Some people may be concerned that the presence of light particles such as $\phi$ may affect cosmology.
However, if the mixing between $\phi$ and the Higgs field is not too small, $\phi$ decays quickly, even if it is generated in the early universe.
Therefore, $\phi$ does not affect the current cosmology scenario.
However, if the mixing of $\phi$ and the Higgs field is not very small, $\phi$ may be found in accelerator experiments

\clearpage

\section{Other possibilities suggested by MPP}
\label{s:mpp}
In this section, we discuss other possibilities than classical conformality that MPP suggests.

\subsection{General cases with $\mathbb{Z}_2$ Symmetry}
In the previous sections, we have considered classical conformality, that is, the case where the renormalized masses are zero. Now let's return to the general case where they are not restricted to zero. 
For simplicity, if we consider mass terms as perturbations for a while, the effective potential is given by Eq.(\ref{e:epotv}) with an additional mass term:
\begin{equation}
\label{e:vemp}
V_{\rm{eff}} = \frac{m^{2}}{2}\phi^{2} + \frac{\kappa^2(\mu_{*})}{128 \pi^2} \phi^{4} \left(\log\qty(\frac{\phi}{v})-\frac{1}{4}\right).
\end{equation}

\begin{figure}[tbp]
\centering
\includegraphics[width=14cm,pagebox=cropbox]{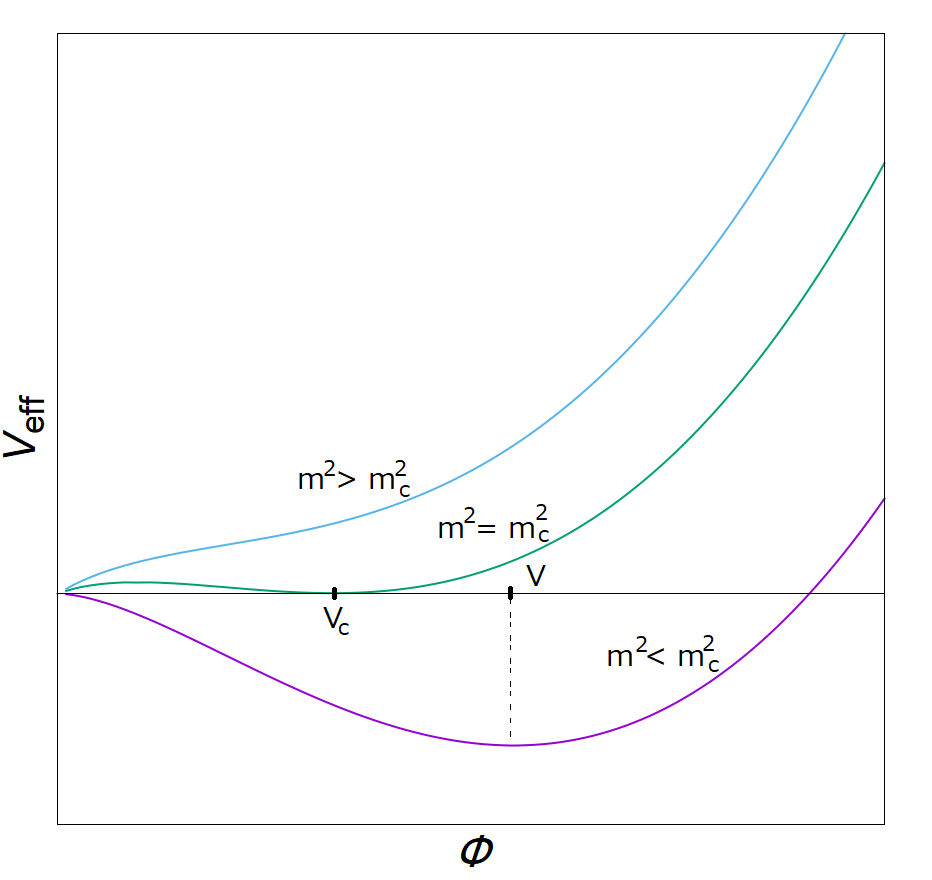}
\caption{the effective potential with the mass term for $m^2<m_c^2\,,\,m^2=m_c^2$ and $m^2>m_c^2$}
\label{f:veff}
\end{figure}

Then, if $ m ^ {2}> 0 $, $ V_ {\rm{eff}} $ generally has two local minima, and if $ m ^ {2} $ is increased from zero, at a certain value $ m ^ {2} = m _ {c} ^ {2} $, the two local minima take the same value (see Fig.\ref{f:veff}).
In other words, when $ m ^ {2} $ is changed as a parameter, the system undergoes a first-order phase transition at $ m ^ {2} = m _ {c} ^ {2} $.


As we have considered in the previous sections, classical conformality can be regarded as a consequence of MPP in a generalized sense, that is, "The coupling constants are fixed at such values that the time evolution of the universe changes drastically when they are changed."
In fact, assuming that the universe starts from the state of $ \langle \phi \rangle = 0 $, if $ m ^ {2} $ is positive then that state is metastable, so the universe remains in that state for a while. However, if $ m ^ {2} $ is negative, that state is unstable, and the universe immediately transitions to another state.
Therefore, the time evolution of the universe changes drastically at the point $ m ^ {2}=0 $.

On the other hand, the original MPP is that "The coupling constants are fixed at such values that the phase of the vacuum changes when they are changed ", and the condition $ m ^ {2} = m _ {c} ^ {2} $ considered above exactly corresponds to this. 

Because the circumstances are the same for $\phi$ and $\varphi$, MPP suggests the following four possibilities for the $Z_{2}\times Z_{2}$ invariant two scalar model:
(1) $ m ^ {2} = 0 $ and ${m'}^ {2} = 0 $, 
(2) $ m ^ {2} = m _ {c} ^ {2}$ and $ {m'}^ {2} = 0 $, 
(3) $ m ^ {2} = 0$ and $ {m'}^ {2} = {m'}^{2} _ {c} $ and
(4) $ m ^ {2} = m _ {c} ^ {2}$ and $ {m'}^ {2} = {m'}^{2}_ {c} $.
Here $m^ {2}$ and $ {m'}^ {2}$ are the squared renormalized masses of $\phi$ and $\varphi $, respectively.

Case (1) is the classical conformality discussed in the previous sections.
 In this case, the effective potential takes two minima at $ (\phi, \varphi) = (v, 0) $ and $ (0, v') $. Which is the true vacuum depends on the magnitude of $ \rho $ and $ \rho'$. 
In (2), the mass of $ \phi $ is chosen to realize the first-order phase transition, while the renormalized mass of $ \varphi $ is chosen to be zero.
In this case again the effective potential takes local minima at two points, $ (v _ {c}, 0) $ and $ (0, v') $. Here, $ v _ {c} $ is the value of $ \phi $ at the first-order phase transition point.
The value of the effective potential at the former is zero, because it is degenerate with the origin $ (0,0) $.
On the other hand, the latter has the same value as the case of classical conformality, which is negative.
Therefore, the latter is the true vacuum, regardless of $ \rho $ and $ \rho'$.
(3) is the case of (2), where $ \phi $ and $ \varphi $ are interchanged.
In (4), both masses are chosen to achieve a first order phase transition. 
In this case, the origin $ (0,0) $ is also a minimum point, and the effective potential takes a minimum value of 0 at three points $ (0,0) $, $ (v _ {c}, 0) $ and $ (0, v'_ {c}) $.

Finally, let's evaluate the renormalized mass squared and the vacuum expectation value of the field at the first-order phase transition point. 
The situation is the same for $\phi $ and $\varphi $, so here we consider the case where $ \phi $ undergoes the first-order phase transition.
As in the previous sections, if $\kappa> 0$, $\varphi $ acquires a positive mass squared from the $ \kappa \phi^{2} \varphi^{2} $ term, so the vacuum expectation value of $\varphi$ is zero. 
Therefore, the effective potential can be considered by introducing mass terms into Eq.(\ref{e:generalve}) and setting $ \varphi = 0 $.
Furthermore, as in the previous sections, if the renormalization point is set to $ \rho(\mu_{*}) = 0 $, then it becomes
\begin{equation}
\label{e:vem}
V_ {\rm{eff}} = \frac{m^{2}}{2}\phi^2 + \frac{1}{64 \pi^2}\qty(\frac{\kappa(\mu_{*})}{2}\phi^2+{m'}^{2})^2 \left (\log \frac{\frac{\kappa(\mu_{*})}{2}\phi^2+{m'}^{2}}{\mu_{*}^2}-\frac{1}{2} \right),
\end{equation}
where $m^2$ and ${m'}^2$ are the renormalized mass squares of $\phi$ and $\varphi$, respectively.

The argument of the second term in this expression is a linear combination of $ \phi^2 $ and $ {m'}^ 2 $.
As checked below, it turns out that the former is sufficiently greater than the latter, if the system is on the first order phase transition point and $\phi$ is near the minimum point, that is, $ m^2 = m_{c}^2 $ and $ \phi \sim v_{c} .$ 
Therefore, the term proportional to $ {m'}^2 $ can be ignored in the argument of the second term, then Eq.(\ref{e:vem}) is reduced to Eq.(\ref{e:vemp}).
By simple calculation, the critical value of $ m^2 $ for Eq.(\ref{e:vemp}) is given by
\begin {equation}
m_{c} ^ 2 = \frac {\kappa^ 2} {128 \pi^{2} \sqrt {e}} v^2,
\end {equation}
and the minimum point is at
\begin {equation}
\phi=v_{c} = \frac{v}{\sqrt {e}}.
\end {equation}

From this result, we can justify what is stated above for the argument of the second term in Eq.(\ref{e:vem}).
In fact,
$ \frac {\kappa} {2} v_ {c} ^ 2: m_ {c} ^ 2 = 1: \frac { \kappa \sqrt {e}} {64 \pi ^ 2}, $
and because the $ {m'}^2 $ term is in the same order as $m_c^2$, it can be ignored compared to the $ \phi^2 $ term as long as $ \kappa $ is perturbative.

Next, let us consider the masses of particles in this vacuum.
Denoting the masses of $\phi$ and $\varphi$ by $\widetilde{M}_{\phi}$
and $\widetilde{M}_{\varphi}$, respectively,
we easily obtain
\begin {eqnarray}
\widetilde{M}_{\phi}^2 = \frac{\mathrm{d}^2}{\mathrm{d} \phi^2} V_{\rm {eff}} 
\eval{}_{v = v_{c}}= \frac{1}{2} M_{\phi}^2, \\
\widetilde{M}_{\varphi}^ 2 = \frac{\kappa}{2} v_ {c} ^ 2 = \frac{1} {e} M_{\varphi} ^ 2,
\end {eqnarray}
where $M_{\phi}$ and $M_{\varphi}$ are the masses of $\phi$ and 
$\varphi$ in the classically conformal vacuum given by Eq.(\ref{e:massphik})
and Eq.(\ref{e:massvphi}), respectively.
From this, it can be seen that, although the classically conformal vacuum and the vacuum at the first order phase transition point are different, the mass scales generated are of the same order of magnitude. In this sense, all the cases (1)-(4) are equally important.

Finally, it should be noted that the observations made here are also valid for the system consisting of the complex scalar field and the gauge field originally considered by Coleman and Weinberg. 
That is, from the point of view of MPP, the first order phase transition point is as interesting as classical conformality.

\subsection{Further Generalization -Cases without $\mathbb{Z}_2$ Symmetry- }
Let's further generalize the argument in the previous subsection and consider the case where there is no $\mathbb{Z}_{2}$ symmetry for $\phi$. However, we assume $\mathbb{Z}_{2}$ invariance for $\varphi$.

Then, the renormalized Lagrangian is as follows:
\begin {equation}
\mathcal{L}=\mbox{(r.h.s. of Eq.(\ref{e:lmdl}) )}+g\phi+\frac{1}{3!}h\phi^3+\frac{\sigma}{2}\phi\varphi^2 .
\end {equation}

\noindent
Here, $g, h, \sigma$ are coupling constants newly introduced by not imposing $Z_{2}$ symmetry on $\phi$. Here, the coefficient of $\phi$ can be taken to be zero, $g=0$, by shifting $\phi$ appropriately. After all, $h$ and $\sigma$ are two new things to consider.

In the previous subsection, we have seen that $m^2$ and $m'^2$ are determined by MPP, but here, we want to determine the four parameters, $m^2$, $m'^2$, $h$ and $\sigma$, by MPP. 
In order to do so, generally, it is sufficient to consider a quadruple critical point, that is, a critical point where the effective potential satisfies four conditions simultaneously.
However, moving four parameters to find critical points is rather cumbersome. Instead, we will construct an example of a critical point that does not have $\mathbb{Z}_{2}$ symmetry.

First, note that the origin is a stationary point of the effective potential, since the renormalized coupling $g$ is 0. Then, as in the previous discussions, it can be seen that the equations $m^2 = 0$ and $m'^2 = 0$ are respectively criticality conditions. That is because if the universe starts from the origin $(\phi, \varphi) = (0, 0)$, its time evolution is greatly different depending on the signs of $m^ 2$ and $m'^ 2$. In the following, we consider the case where $m^ 2 = m'^ 2 = 0$ holds.

Then the remaining parameters are $h$ and $\sigma$.
Considering the behavior near the origin $(\phi, \varphi) = (0, 0)$, it can be seen that the equation $\sigma = 0$ is also a criticality condition under the assumption of $m^ 2 = m'^ 2 = 0$. 
 In fact, the behavior of the effective potential near the origin
$V_{\rm {eff}} \sim \frac {h} {6} \phi ^ 3 + \frac {\sigma} {2} \phi \varphi ^ 2 $
changes largely depending on the signs of $h$ and $\sigma$. In the following, we will concentrate on the case where $\sigma = 0$.

Then, the only remaining parameter is $h$, and in order to determine it, it is sufficient to consider the ordinary critical point of the effective potential. 
If the renormalization point is chosen so that $\rho(\mu_ {*})=0$ as before, and the vacuum expectation value of $\varphi$ is 0, the effective potential can be approximated by the following equation:
\begin{equation}
V_ {\rm{eff}} = \frac{h}{6}\phi^3+ \frac{\kappa^2}{256 \pi^2} \phi^{4} \left (\log \frac{\phi^2}{v^2}-\frac{1}{2} \right). 
\end{equation}

\begin{figure}[tbp]
\centering
\includegraphics[width=0.7\textwidth,pagebox=cropbox]{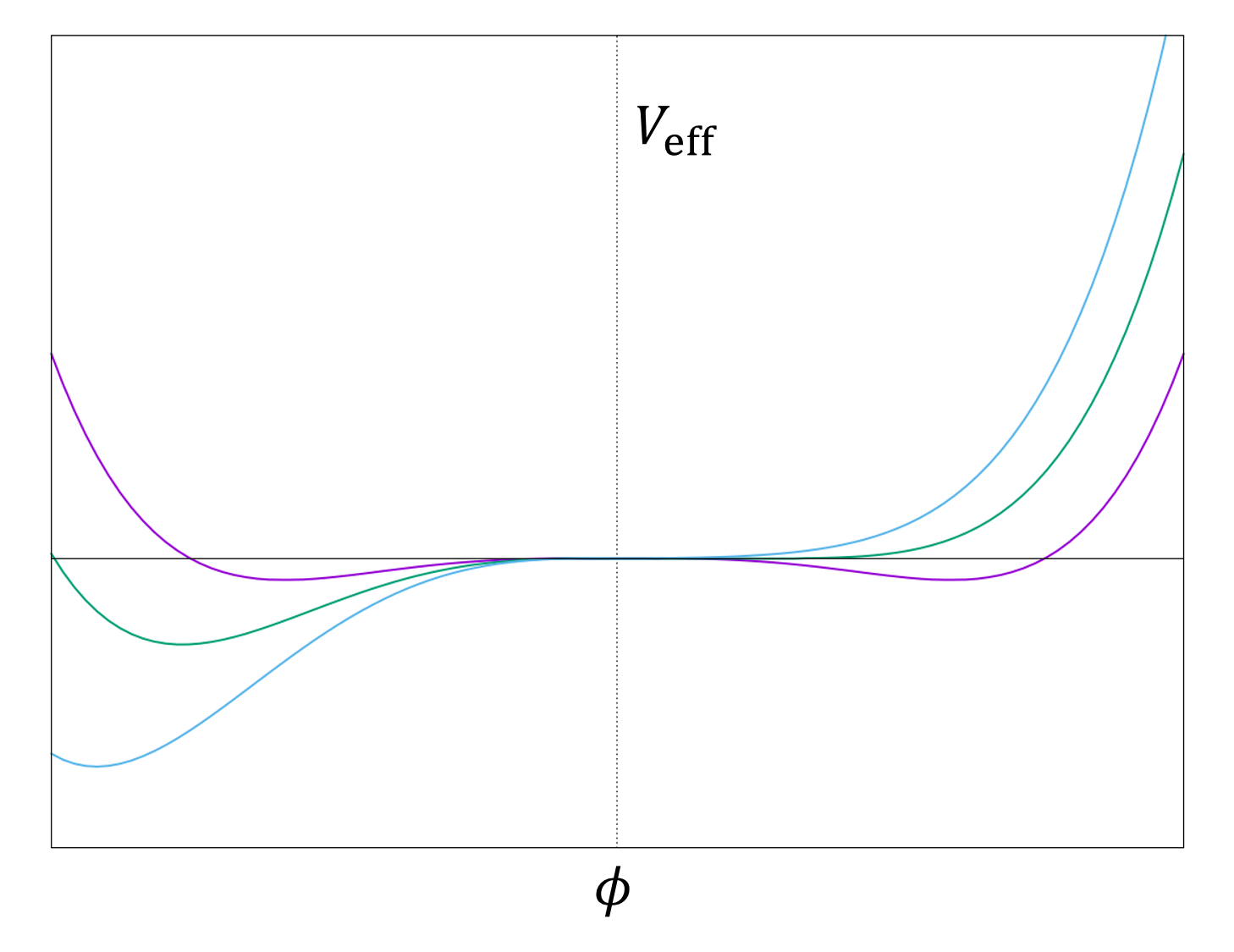}
\caption{The effective potential with $h$ increased from 0 (purple$\to$green$\to$blue).}
\label{f:hinc}
\end{figure}

Fig.\ref{f:hinc} shows how the effective potential changes when changing $h$. This is similar to the previous subsection, but here we consider $\phi^3 $ as a perturbation rather than $\phi^ 2 $. Therefore, there is no $ \mathbb{Z}_{2} $ symmetry except when $ h = 0 $. In the following, it is sufficient to consider the case of $ h \geq 0 $, because if we redefine $ \phi $ as $-\phi $, the sign of $ h $ will change.

Let's see what happens when we increase $ h $ from $ 0 $. 
First, in the case of $ h = 0 $, it is the classically conformal case, and the effective potential vanishes up to the third derivative at the origin. 
Therefore, the criticality at the origin is higher by one than that of $ h \neq 0 $, and should be adopted from the viewpoint of MPP. 
In this case, the effective potential has a double well shape and is symmetrical.

\begin{figure}[tbp]
\begin{tabular}{cc}
\begin{minipage}[t]{0.45\hsize}
\centering
   \includegraphics[width=\textwidth]{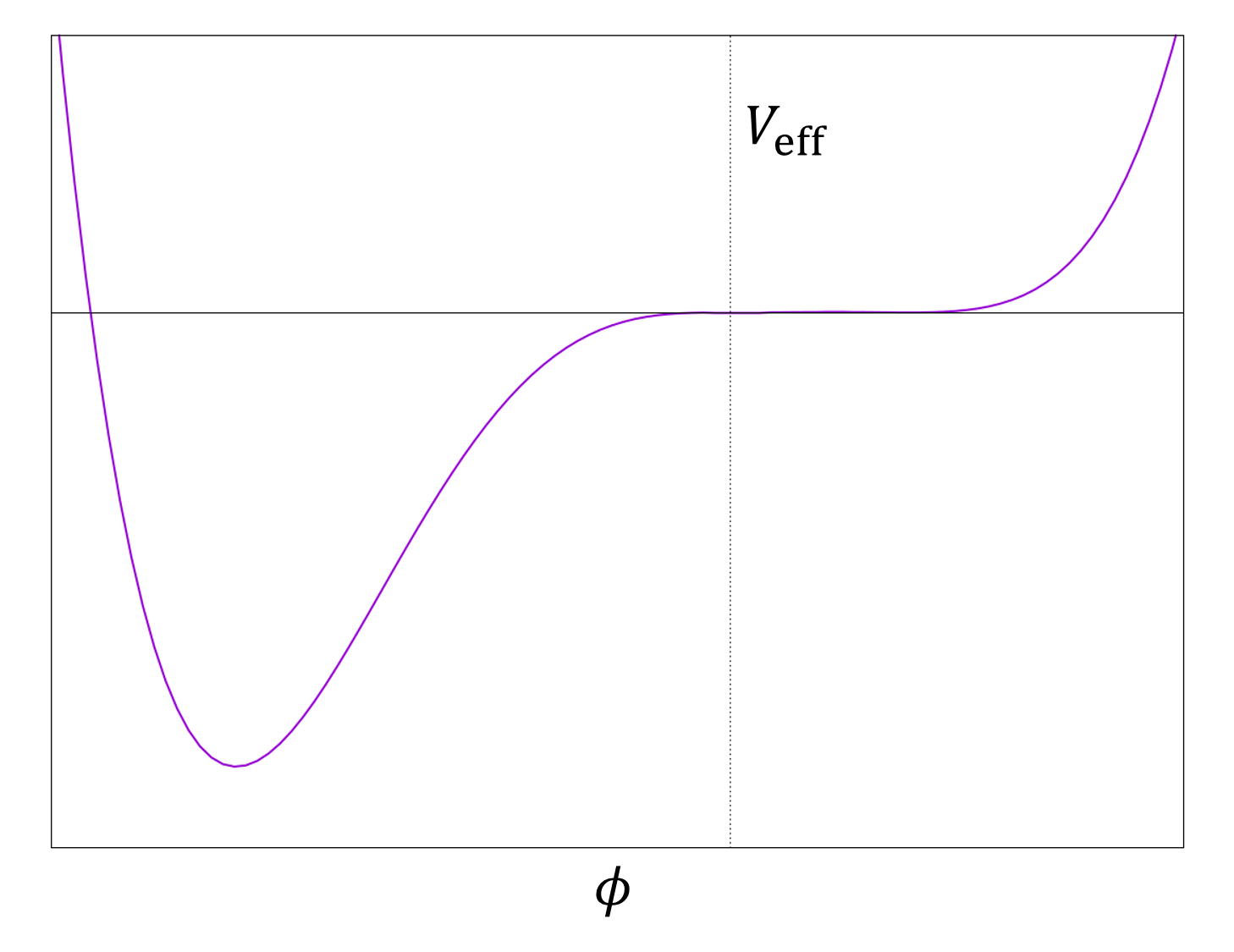}
  \subcaption{The overall view.}

\end{minipage}
&
\begin{minipage}[t]{0.45\hsize}
\centering
   \includegraphics[width=\textwidth]{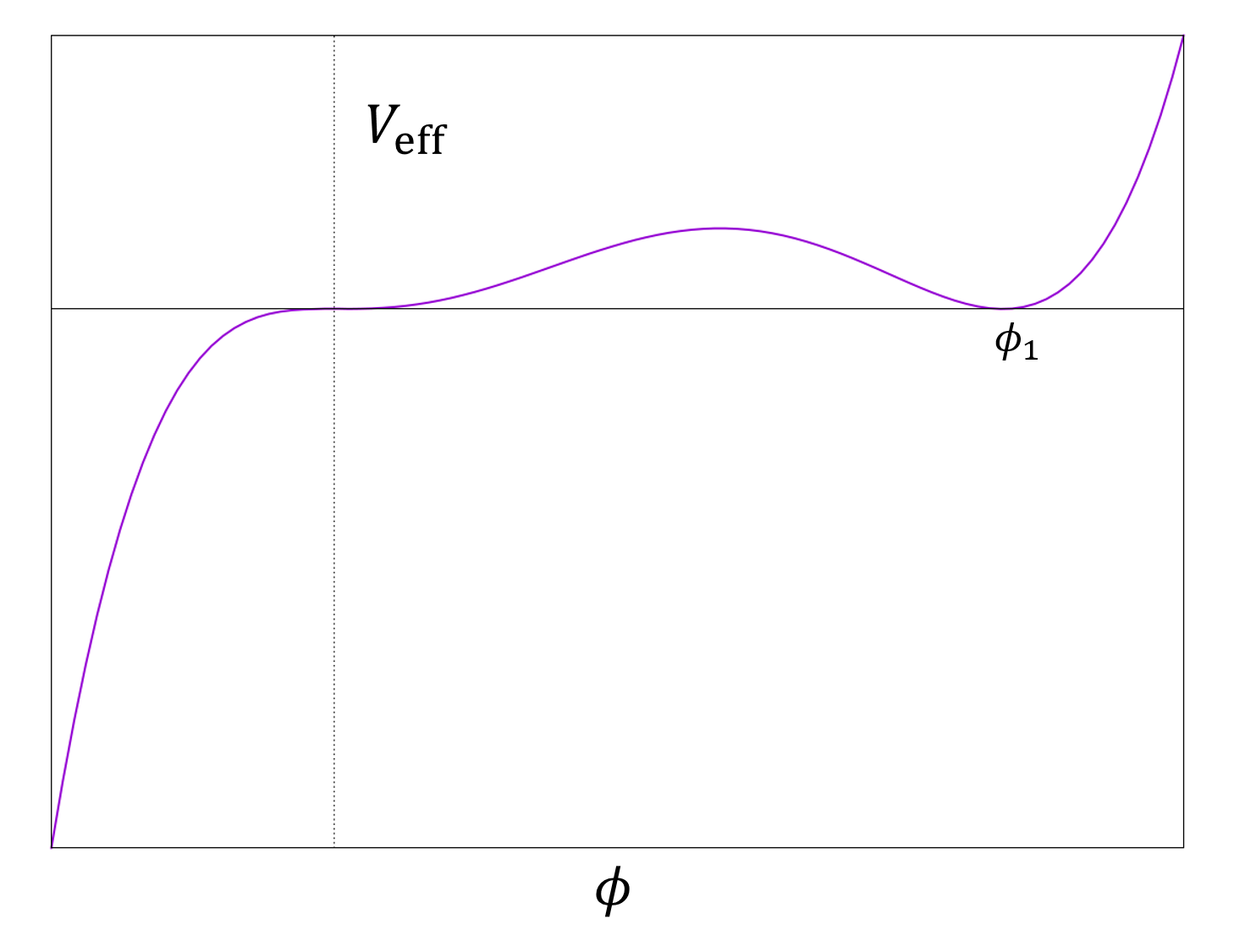}
  \subcaption{The vicinity of the origin.}
  \label{f:vich1}
  \end{minipage}
\end{tabular}
\caption{The effective potential at $h=h_1$.}
  \label{f:h1}
\end{figure}

As we increase $ h $, the left well gets deeper and the right well gets shallower (Fig.\ref{f:hinc}). 
When $  h = h_ {1} $, the value at the bottom of the right well at $ \phi = \phi_ {1} $ becomes zero and degenerates to that at the origin $ \phi = 0 $
(see Fig.\ref{f:h1}). This point is to be adopted from the view point of MPP as well as the first-order phase transition point discussed in the previous subsection.

\begin{figure}[tbp]
\begin{tabular}{cc}
\begin{minipage}[t]{0.45\hsize}
\centering
   \includegraphics[width=\textwidth]{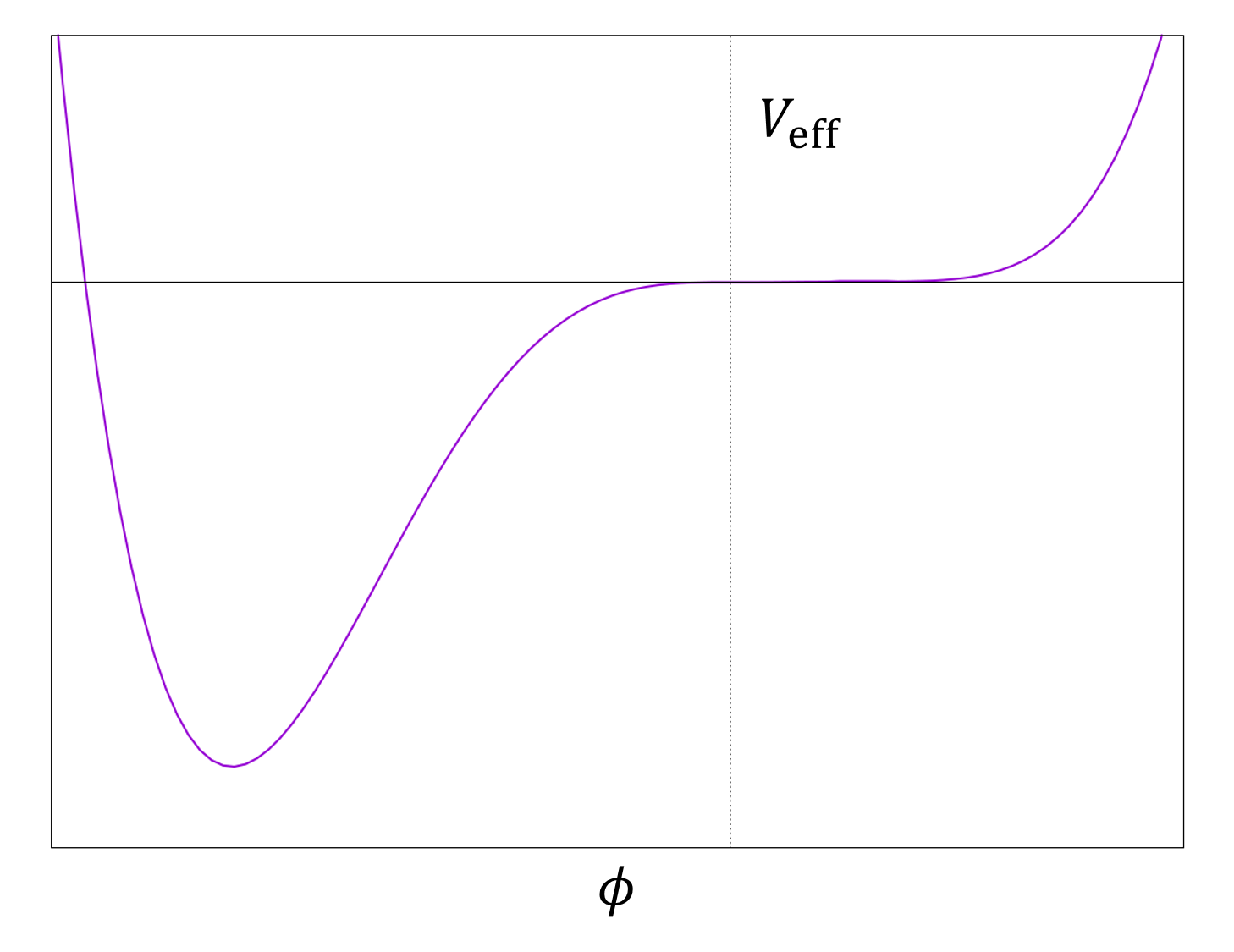}
  \subcaption{The overall view.}
\end{minipage}
&
\begin{minipage}[t]{0.45\hsize}
\centering
   \includegraphics[width=\textwidth]{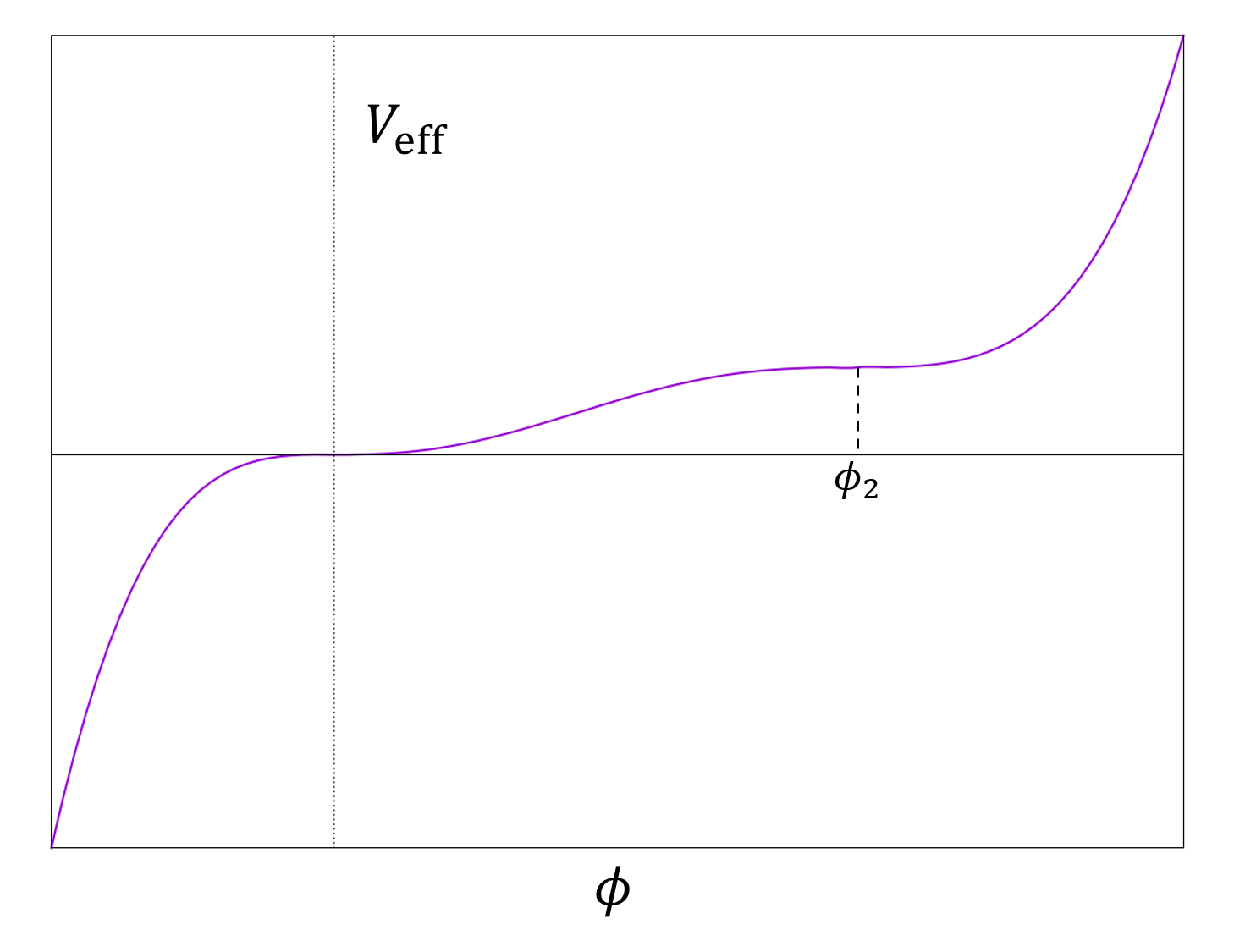}
  \subcaption{The vicinity of the origin.}
  \label{f:vich2}
\end{minipage}
\end{tabular}
\caption{The effective potential at $h=h_2$.}
  \label{f:h2}
\end{figure}

When $ h $ is further increased, when $ h = h_{2} $,  the right well becomes a saddle point and disappears (see Fig.\ref{f:h2}).
We denote the position of the saddle point at that time as $ \phi = \phi_{2} $.
Again, the criticality of the effective potential has increased, so it should be adopted from the point of view of MPP.
Even if we increase $h$ any more, the behavior of the effective potential does not change and the criticality is not enhanced.

In fact, we can confirm the above statements by a simple numerical calculation, and obtain the following values;
\begin {equation}
h_ {1} = 0.70855 \frac{\kappa ^ 2 v} {32 \pi ^ 2},\ 
\phi_ {1} = 0.4723 v,\ 
h_ {2} = 0.73576 \frac{\kappa ^ 2 v} {32 \pi ^ 2},\ 
\phi_ {2} = 0.3679 v.
\end {equation}

As in the previous subsection, in these cases also, the mass scale produced is comparable to $ v $, which is as interesting as the classically conformal case. In particular, neither of the two critical points discussed here has $ \mathbb{Z}_{2} $ symmetry for $ \phi $.
Therefore, even if $ \phi $ has a vacuum expectation value, it does not cause the cosmological domain wall problem, and it can be used to construct more realistic models.

\newpage
\section{Conclusion}
In this paper, we have examined the possibility that the weak scale is generated dynamically from the Planck scale.
In particular, we have considered the classically conformal $\mathbb{Z}_2\times \mathbb{Z}_2$ invariant two scalar model as a minimal model in which the mass scale is generated nonperturbly from the cutoff.
We have also investigated whether it is possible to reproduce the mass term and vacuum expectation value of the Higgs field by coupling this model with the standard model in the Higgs portal framework.
There are two main results.

The first is that only one of $\mathbb{Z}_2\times \mathbb{Z}_2$ symmetry is spontaneously broken in  the classically conformal $\mathbb{Z}_2\times \mathbb{Z}_2$ invariant two scalar model.
The relationship between the vacuum expectation value \m \expval{\phi}\m\, and the scale \m\mu_*\m\, where \m\rho(\mu_*)=0\m\, is given by 
\begin{align*}
\tag{\ref{e:vev}}
\expval{\phi}^2=\f{2}{\kappa(\mu_*)}\mu_*^2.
\end{align*}
Also in a special case, we get explicit relationship between the cutoff and the vacuum expectation value:
\begin{align*}
\tag{\ref{e:expvphi}}
\expval{\phi}^2\simeq\Lambda^2\dfrac{2}{\kappa_0}\exp\qty(-\dfrac{32\pi^2}{3} \dfrac{\rho_0}{\kappa_0^2}).
\end{align*}

These results are obtained only from the assumption that $\rho(\mu)$ becomes 0 first among  the running coupling constants when lowering the renormalization scale.
Because this assumption holds in a wide range of initial values of $\rho_0,\rho'_0$ and $\kappa_0$, this mechanism is universal.

Secondly, by coupling this model to the standard model in the Higgs portal framework, this mechanism can be used to generate the weak scale from the Planck scale.
Furthermore, we show that the scalar field $\varphi$ without vacuum expectation value must have a mass greater than 0.6 TeV, and that the scalar field $\phi$ with vacuum expectation value must have a smaller mass.
We also have confirmed that the mixing angle between the Higg field and the scalar field with vacuum expectation value is small enough not to be excluded experimentally.
We emphasize that the scalar field $\varphi$ without vacuum expectation value can be regarded as dark matter.

On the other hand, the mass of the Higgs portal dark matter has a lower bound of 0.7 TeV from the direct search experiment of dark matter.
Furthermore, assuming Higgs inflation, the tensor to scalar ratio of the cosmic microwave background gives its mass an upper bound of 1.2 TeV.
These are consistent with our analysis that the mass is over 0.6 TeV.

Considering $\phi$ in turn, the important point is that it mixes with the Higgs field.
If the mixing is not so large, there is no contradiction with the accelerator experiments.
If the mixing is not so small, $\phi$ does not affect cosmology because it quickly decays after being generated in the early universe.
Furthermore, if the mixing angle is not too small, it will be detected in near future accelerator experiments such as precise measurements of the Higgs' decay\footnote{
We plan to post a paper that examines these in detail in the near future.
}.

In this paper, with the exception of Section \ref{s:mpp}, we have focused on classical conformality.
We have further pointed out that classical conformality can be understood from MPP.
When going back to the original MPP, there is a possibility that, besides classical conformality, the parameters of the theory are chosen to be the first order phase transition point, which is  discussed in Section 6.
In the cases of the classical conformality and the first-order phase transition point, the vacua are different but the mass scales are of  similar size.
Therefore, they are equally interesting.
Even in the conventional Coleman-Weinberg mechanism, it is interesting to consider the first-order phase transition point instead of classical conformality.
Besides these, as mass scales other than the Planck scale, there are the Majorana mass and the cosmological constant as well as the weak scale.
It would be interesting to construct a minimal model to explain them based on MPP.

\section*{acknowledgment}
We thank Koji Tsumura and Masatoshi Yamada for useful discussions, and Kin-ya Oda and Yuta Hamada for helpful comments.

\newpage

\appendix

\renewcommand{\theequation}{A.\arabic{equation}}
\setcounter{equation}{0}
\section{Calculation of effective potential in the large N limit}
\label{s:calclnve}
We explain here the detailed calculation of the effective potential of $O(N)\times O(N)$ scalar model in the large N limit.
The bare Lagrangian is
\begin{multline}
\mathcal{L}
=N\left(\frac{1}{2}(\del_\mu\phi_i)^2+
\frac{1}{2}(\del_\mu\varphi_i)^2+\right.\\
\left.\frac{1}{2}(m_0^2\phi_i^2+{m'}_0^{2}\varphi^2)+
\frac{\rho_0}{8}(\phi_i^2)^2+
\frac{\kappa_0}{4}(\phi_i^2)(\varphi_j^2)+
\frac{\rho'_0}{8}(\varphi_i^2)^2\right),
\end{multline}
where $\phi_i,\varphi_i \in \mathbb{R},\,i=1,\ldots,N,\,\phi_i^2=\sum_i \phi_i\phi_i,\,\varphi_i^2=\sum_i \varphi_i\varphi_i$, and $m_0,m'_0,\rho_0,\rho'_0$ and $\kappa_0$ are bare couplings.
Here we introduce
\begin{align}
\lambda_0:=\mqty(\rho_0&\kappa_0\\\kappa_0&\rho'_0),
\end{align}
then $\mathcal{L}$ can be rewritten as
\begin{align}
\a\mathcal{L}\\
\a=N
\left(\frac{1}{2}(\del_\mu\phi_i)^2+
\frac{1}{2}(\del_\mu\varphi_i)^2+
\frac{1}{2}(m_0^2\phi_i^2+{m'}_0^{2}\varphi^2)+
\f{1}{8}\mqty(\phi_i^2\a\varphi_j^2)\lambda_0\mqty(\phi_i^2\\\varphi_j^2)
\right),
\\
\a=N
\left(\frac{1}{2}(\del_\mu\phi_i)^2
+\frac{1}{2}(\del_\mu\varphi_i)^2
+\qty(\f{\lambda_0}{2}\mqty(\phi_i^2\\\varphi_j^2)+\mqty(m_0^2\\{m'}_0^{2}))^{t}
\f{\lambda_0^{-1}}{2}
\qty(\f{\lambda_0}{2}\mqty(\phi_i^2\\\varphi_j^2)+\mqty(m_0^2\\{m'}_0^{2}))
\right)+(const.).
\end{align}

Let us calculate the effective potential.
The partition function $Z$ is defined by
\begin{align*}
Z:=\int D\phi D\varphi\exp\qty(-\int d^4x \qty[\mathcal{L}+N\qty(J_{1i}\phi_i+J_{2i}\varphi_i)]),
\end{align*}
where $J_{1i}$ and $J_{2i}$ are the source fields.

First we 
rewrite $Z$ with the auxiliary fields $C:=\mqty(c_1\a c_2)^t$ as 
\begin{multline}
 Z \propto\int D\phi D\varphi DC
\,\exp\left(-N\int d^4x \left[\frac{1}{2}\qty((\del_\mu\phi_i)^2
+(\del_\mu\varphi_i)^2)
-\f{1}{2}C^{t}\lambda_0^{-1}C\right.\right.
\\\left.\left.+C^{t}\lambda_0^{-1}
\qty(\f{\lambda_0}{2}\mqty(\phi_i^2\\\varphi_j^2)+\mqty(m_0^2\\{m'}_0^{2}))
+J_{1i}\,\phi_i+J_{2i}\,\varphi_i\right]\right).
\end{multline}
Then, integarting $\phi_i$, we get
\begin{align}
\notag\a Z[J]\\\notag
\a\propto\int DC
\exp\left(-\f{N}{2}\tr\log(-\del^2+\hat{C})-\right.
\\\a\qquad\qquad\qquad\qquad
\left.\int d^4x \,N\qty[
-\f{1}{2}C^{t}\lambda_0^{-1}C
+C^{t}
\lambda_0^{-1}\mqty(m_0^2\\{m'}_0^{2})]
+\f{N}{2}J^{t}((-\del^2+\hat{C})^{-1}\otimes I_N )\,J\right),
\end{align}
where $J:=\mqty(J_{1i}&J_{2i})^{t} \,,\, \hat{C}=\mqty(c & 0 \\ 0 & c' )$ and $I_N$ is the unit matrix of size $N$.
Because $C$ integral is equivalent to substituting the value of the stationary point for $C$ in the large N limit, $Z$ becomes
\begin{align}
Z=\exp (-W[J,C[J]]),
\end{align}
where we have defined
\begin{multline}
\f{W[J,C]}{N}=\f{1}{2}\tr\log(-\del^2+\hat{C})+
\\
\int d^4x \qty[
-\f{1}{2}C^{t}\lambda_0^{-1}C
+C^{t}\lambda_0^{-1}\mqty(m_0^2\\{m'}_0^{2})]
-\f{1}{2}J^{t}((-\del^2+\hat{C})^{-1}\otimes I_N )\,J.
\end{multline}
Here $C[J]$ is determined from
\begin{align}
\fdv{C}W[J,C]\eval{}_{C=C[J]}=0.
\end{align}

Then the effective action \m \Gamma[\phi_i,\varphi_i]\m\, is given by the Legendre transformaiton of \m W[J]\m:
\begin{align}
\f{\Gamma[\phi_i,\varphi_i]}{N}
:=\min_J \qty(\f{W[J,C[J]]}{N}-\mqty(\phi_i\a\varphi_i)J),
\end{align}
that is, $\Gamma$ is give by
\begin{align}
\label{e:defgam}
\f{\Gamma[\phi_i,\varphi_i]}{N}=\qty(\f{W[J,C[J]]}{N}-\mqty(\phi_i\a\varphi_i)J)\eval{}_{J=J(\phi_i,\varphi_i)},
\end{align}
where $J(\phi_i,\varphi_i)$ is the solution of
\begin{align}
\label{e:detJ}
0=\fdv{J}\qty(\f{W[J,C[J]]}{N}-\mqty(\phi_i\a\varphi_i)J)\eval{}_{J=J(\phi_i,\varphi_i)}=\f{1}{N}\fdv{J}W[J,C[J]]-\mqty(\phi_i\a\varphi_i)\eval{}_{J=J(\phi_i,\varphi_i)}.
\end{align}
However, from the definition of $C[J]$,
\begin{align}
\fdv{C}W[J,C]\eval{}_{C=C[J]}=0,
\end{align}
for any $J$.
Therefore
\begin{align}
\fdv{J} W[J,C[J]]=\qty(\fdv{C}{J}\fdv{C}+\widetilde{\fdv{J}})W[J,C[J]]=\widetilde{\fdv{J}}W[J,C[J]],
\end{align}
where $\widetilde{\fdv{J}}$ means to differentiate the part that depends on $J$ explicitly.
Then, Eq.(\ref{e:detJ}) becomes
\begin{align}
0\a=\f{1}{N}\fdv{J} W[J,C[J]]-\mqty(\phi_i\a\varphi_i)\eval{}_{J=J(\phi_i,\varphi_i)}=\f{1}{N}\widetilde{\fdv{J}} W[J,C[J]]-\mqty(\phi_i\a\varphi_i)\eval{}_{J=J(\phi_i,\varphi_i)}
\\
\a=\qty(-((-\del^2+\hat{C})^{-1}\otimes I_N )\,J-\mqty(\phi_i\\\varphi_i))^{t}\eval{}_{J=J(\phi_i,\varphi_i)},
\end{align}
which gives
\begin{align}
J(\phi_i,\varphi_i)=-((-\del^2+\hat{C})\otimes I_N )\,\mqty(\phi_i\\\varphi_i).
\end{align}
Substituting this into Eq.(\ref{e:defgam}), we get
\begin{multline}
\f{\Gamma}{N}= \f{1}{2}\tr\log(-\del^2+\hat{C})+\int d^4x \qty[
-\f{1}{2}C^{t}\lambda_0^{-1}C
+C^{t}\lambda_0^{-1}\mqty(m_0^2\\{m'}_0^{2})]
\\
+\f{1}{2}\mqty(\phi_i\a\varphi_i)((-\del^2+\hat{C})\otimes I_N )\,\mqty(\phi_i\\\varphi_i).
\end{multline}
If $\phi_i\,,\varphi_i$ and $C$  do not depend on $x$,
\begin{align}
\mqty(\phi_i\a\varphi_i)((-\del^2+\hat{C})\otimes I_N )\,\mqty(\phi_i\\\varphi_i)=\int d^4x\, C^{t}\mqty(\phi_i^2\\\varphi_i^2),
\end{align}
and 
\begin{align}
\label{e:trlog}
\tr \log(-\partial^2+\hat{C})
\a=\sum_{a=1,2} \int d^4x \int \f{d^4k}{(2\pi)^4}\, \log(k^2+c_a)\\
\a=\sum_{a=1,2} \int \f{d^4x}{32\pi^2}\qty[(k^4-c_a^2)\log(k^2+c_a)-\f{1}{2}k^4+c_ak^2],
\end{align}
where $c_1:=c\,,\,c_2:=c'$.
Therefore we obtain
\begin{multline}
\f{\Gamma}{N}
= \int d^4x \left\{\sum_{a=1,2} \f{1}{64\pi^2}\qty[(k^4-c_a^2)\log(k^2+c_a)-\f{1}{2}k^4+c_ik^2]^{k=\Lambda}_{k=0}\right.
\\\left.
-\f{1}{2}C^{t}\lambda_0^{-1}C
+C^{t}\lambda_0^{-1}\mqty(m_0^2\\{m'}_0^{2})
+\f{1}{2}C^t\mqty(\phi^2\\\varphi^2)\right\},
\end{multline}
where $\Lambda$ is the momentum cutoff and we also have defined $\phi^2=\phi_i^2$ and $\varphi^2=\varphi_i^2$.
Then the effective potential is 
\begin{align}
\a\f{\Ve}{N}\notag
\\\a=
\sum_{a=1,2} \f{1}{64\pi^2}\qty[(k^4-c_a^2)\log(k^2+c_a)
-\f{1}{2}k^4+c_ak^2]^{k=\Lambda}_{k=0}
-\f{1}{2}C^{t}\lambda_0^{-1}C
+C^{t}\lambda_0^{-1}\mqty(m_0^2\\{m'}_0^{2})
+\f{1}{2}C^t\mqty(\phi^2\\\varphi^2)
\\\a=
\sum_{a=1,2} \f{1}{64\pi^2}\qty[(\Lambda^4-c_a^2)\log(\Lambda^2+c_a)+c_a^2\log(c_a)+2c_a\Lambda^2]
-\f{1}{2}C^{t}\lambda_0^{-1}C
+C^{t}\lambda_0^{-1}\mqty(m_0^2\\{m'}_0^{2})
+\f{1}{2}C^t\mqty(\phi^2\\\varphi^2).
\end{align}
Here we have dropped the constant terms.
Furthermore, ignoring the terms that disappear with $\Lambda\to\infty$, we have
\begin{align}
\frac{\Ve}{N}=
\sum_{a=1,2}\frac{c_a^2}{64\pi^2}\qty(\log\frac{c_a}{\Lambda^2}-\frac{1}{2})
-\frac{1}{2} C^{t}\lambda_0^{-1} C
+C^{t}\qty(\lambda_0^{-1}\mqty(m_0^2\\{m'}_0^{2})+\f{\Lambda^2}{32\pi^2}\mqty(1\\1))
+\frac{1}{2}C^{t}\mqty(\phi^2 \\ \varphi^2).
\end{align}
Then we define the renormalized couplings  $m^2(\mu),{m'}^2(\mu)$ and $\lambda(\mu)=\mqty(\rho(\mu)\a\kappa(\mu)\\\kappa(\mu)\a\rho'(\mu))$ as
\begin{align}
\lambda_0^{-1}\mqty(m_0^2\\ {m'}_0^{2})+\frac{\Lambda^2}{32\pi^2}\mqty(1\\1)\a=\lambda^{-1}(\mu)\mqty(m^2(\mu)\\ {m'}^{2}(\mu)),
\\
-\lambda_0^{-1}+\frac{1}{32\pi^2}\log\qty(\f{\mu^2}{\Lambda^2})\mqty(1\a0\\0\a1)\a=-\lambda^{-1}(\mu),
\end{align}
and the final expression of the effective potential is
\begin{equation}
\frac{\Ve}{N}=-\frac{1}{2} C^{t}\lambda^{-1} C
+\frac{c^2}{64\pi^2}\qty(\log\frac{c}{\mu^2}-\frac{1}{2})
+\frac{{c'}^2}{64\pi^2}\qty(\log\frac{{c'}}{\mu^2}-\frac{1}{2})
+C^{t}\lambda^{-1}\mqty(m^2\\{m'}^{2})
+\frac{1}{2}C^{t}\mqty(\phi^2 \\ \varphi^2).
\end{equation}
If \m m(\mu)={m'}(\mu)=0\m, this equation becomes Eq.(\ref{e:veln}).

\newpage

\renewcommand{\theequation}{B.\arabic{equation}}
\setcounter{equation}{0}
\section{Detailed analysis of large N effective potential}
In this Section, we examine the global structure of the effective potential calculated in Section \ref{s:velnv}.
The effective potential is given by 
\begin{align*}
\tag{\ref{e:veln}}
\frac{\Ve}{N}=-\frac{1}{2} C^{t}\lambda^{-1} C
+\frac{c^2}{64\pi^2}\qty(\log\frac{c}{\mu^2}-\frac{1}{2})
+\frac{c'^2}{64\pi^2}\qty(\log\frac{c'}{\mu^2}-\frac{1}{2})
+\frac{1}{2}C^{t}\mqty(\phi^2 \\ \varphi^2),
\end{align*}
where the auxiliary fields $C=\mqty(c&c')^t$ are determined by
\begin{equation*}
\begin{cases}
\tag{\ref{e:Ccond}}
\displaystyle
\frac{1}{2}\phi^2=\bar{\kappa}c'-\frac{c}{32\pi^2}\log\qty(\frac{c}{\mu_{\rho'}^2}),\\
\displaystyle
\frac{1}{2}\varphi^2=\bar{\kappa}c-\frac{c'}{32\pi^2}\log\qty(\frac{c'}{\mu_{\rho}^2}).
\end{cases}
\end{equation*}
From these, $\Ve\,,\phi$ and $\varphi$ can be considered to be parameterized by c and $c'$.

Let us consider the constraints for $c$ and $c'$.
Because $c$ and $c'$ have the meaning of the mass squares, the sufficient and necessary condition for the vacuum to be stable is that $c$ and $c'$ are nonnegative.
In addition, because $\phi$ and $\varphi$ are real numbers, $\phi^2$ and $\varphi^2$ are nonnegative.

\begin{figure}[htbp]
\centering
\includegraphics[width=0.7\textwidth,pagebox=cropbox]{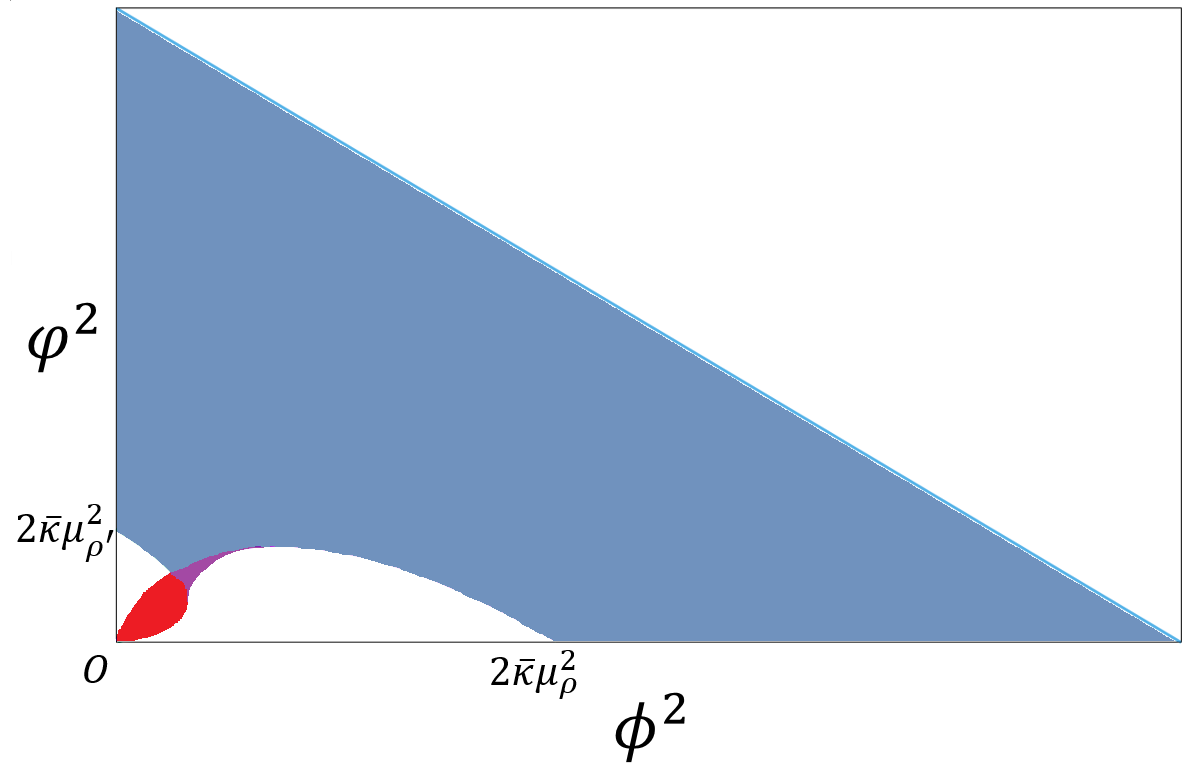}
\caption{$\phi-\varphi$ plane for $c\geq0$ and $c'\geq0$}
\label{f:pv}
\end{figure}

The map $(c,c')\to(\phi,\varphi)$ is illustrate as Fig.\ref{f:pv}.

\begin{figure}[tbp]
\centering
\includegraphics[width=0.7\textwidth,pagebox=cropbox]{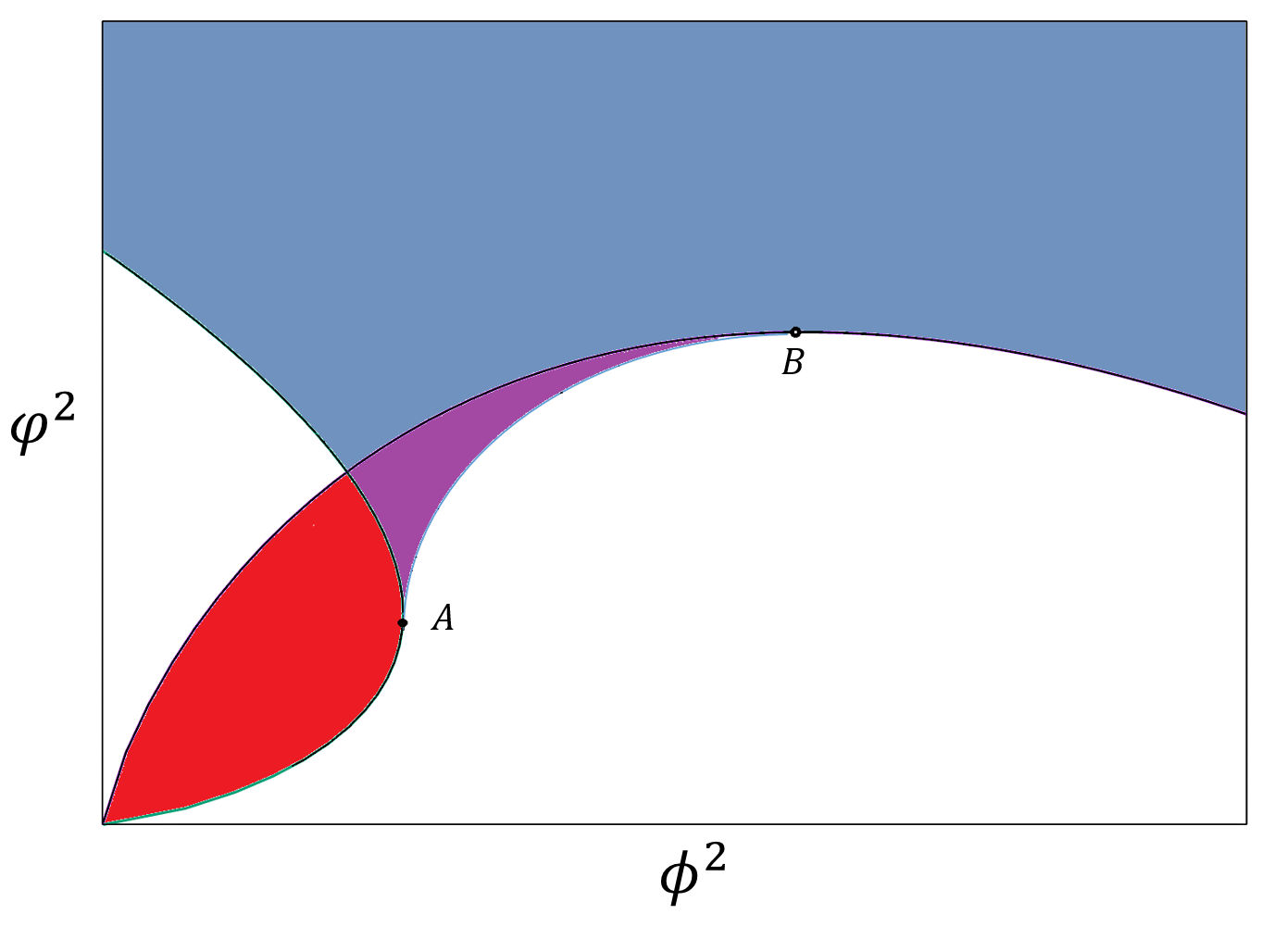}
\caption{The vicinity of the origin of Fig.\ref{f:pv}. The red  and blue areas are pasted along curved line segment AB.}
\label{f:pv2}
\end{figure}

The region where $c$ and $c'$ are large corresponds to the Landau pole, which are ignored in this paper, as the case of (4) in Section \ref{s:ln}.
Fig.\ref{f:pv2} is an enlarged view of the vicinity of the origin of Fig.\ref{f:pv}.

In addition, there is a region (the purple region in Fig.\ref{f:pv} and Fig.\ref{f:pv2}) in the $\phi-\varphi$ plane that is covered twice by the $c-c'$ plane.
Therefore, although $\Ve$ is a single-valued function for $c$ and $c'$, it is a partially bivalent function for $\phi$ and $\varphi$.
It may be necessary to investigate this region in detail.

\clearpage
\bibliography{20190420}
\end{document}